\documentclass[conference]{IEEEtran}
\IEEEoverridecommandlockouts
\usepackage{amsmath}
\usepackage{amsfonts}
\usepackage{bbding}
\usepackage{amssymb}
\usepackage{array}
\usepackage{subfigure}

\usepackage{graphicx}
\usepackage{subfigure}
\usepackage[named]{algo}
\usepackage{algorithmic}
\usepackage{psfrag}
\usepackage{stfloats}
\usepackage[compress]{cite}
\makeatletter
\renewcommand{\citepunct}{,\penalty\@m\hskip.13emplus.1emminus.1em}
\renewcommand{\citedash}{\hbox{--}\penalty\@m}
\makeatother
\usepackage{setspace}
\usepackage{color}
\allowdisplaybreaks

\usepackage{amsthm}
\usepackage{stfloats}

\newtheorem{rem}{Remark}
\newtheorem{pro}{Property}
\newtheorem{prop}{Proposition}

\usepackage{hyperref}
\usepackage{bm}
\renewcommand{\IEEEQED}{\IEEEQEDopen}

\begin{document}
\title{Cross-layer Design for Mission-Critical IoT in Mobile Edge Computing Systems}

\author{Changyang She, Yifan Duan, Guodong Zhao, Tony~Q.~S.~Quek, Yonghui~Li, and Branka~Vucetic,

\thanks{Part of this paper was presented at the International Conference on
Wireless Communications and Signal Processing 2018 \cite{Yifan2018delay}.}

\thanks{C. She, Y. Li and B. Vucetic are with the School of Electrical and Information Engineering, University of Sydney, Sydney, NSW 2006, Australia (email:shechangyang@gmail.com, \{yonghui.li,branka.vucetic\}@sydney.edu.au).}

\thanks{Y. Duan is with the School of Information and Communication Engineering, University of Electronic Science and Technology of China, Chengdu 611731, China (e-mail: duanyifan23@outlook.com).}

\thanks{G. Zhao is with School of Engineering, University of Glasgow, Glasgow, G12 8LT, UK. (e-mail: Guodong.Zhao@glasgow.ac.uk).}

\thanks{T. Q. S. Quek is with the Information Systems Technology and Design Pillar, Singapore University of Technology and Design, 8 Somapah Road, Singapore 487372 (e-mail: tonyquek@sutd.edu.sg). }
}
\maketitle



%
%

\maketitle
\begin{abstract}
In this work, we propose a cross-layer framework for optimizing user association, packet offloading rates, and bandwidth allocation for Mission-Critical Internet-of-Things (MC-IoT) services with short packets in Mobile Edge Computing (MEC) systems, where enhanced Mobile BroadBand (eMBB) services with long packets are considered as background services. To reduce communication delay, the 5th generation new radio is adopted in radio access networks. To avoid long queueing delay for short packets from MC-IoT, Processor-Sharing (PS) servers are deployed at MEC systems, where the service rate of the server is equally allocated to all the packets in the buffer. We derive the distribution of latency experienced by short packets in closed-form, and minimize the overall packet loss probability subject to the end-to-end delay requirement. To solve the non-convex optimization problem, we propose an algorithm that converges to a near optimal solution when the throughput of eMBB services is much higher than MC-IoT services, and extend it into more general scenarios. Furthermore, we derive the optimal solutions in two asymptotic cases: communication or computing is the bottleneck of reliability. Simulation and numerical results validate our analysis and show that the PS server outperforms first-come-first-serve servers.
\end{abstract}

\vspace{-0.2cm}
\section{Introduction}
Mission-Critical Internet-of-Things (MC-IoT) will be widely deployed in future wireless networks for remote health monitoring, haptic interaction, and factory automation \cite{markakis2017efficient,Factory2015Yilmaz}. Achieving ultra-reliable and low-latency communications (URLLC) (e.g., $10^{-7}$ packet loss probability and $1$~ms End-to-End (E2E) delay) for MC-IoT has been considered as one of the major goals in 5th Generation (5G) cellular networks \cite{LatencyPhi}. Most existing technologies mainly focus on one of the seven layers of the open systems interconnection model, and cannot guarantee the E2E delay \cite{Jiang2019Low}. To satisfy the requirements of MC-IoT, we need to re-design the physical-layer resource management, the link-layer scheduling policy, and the network-layer user association from a cross-layer perspective.

One of the major differences between MC-IoT and enhanced Mobile BroadBand (eMBB) services lies in the sizes of packets. With high data rate, the packet size in eMBB services is relatively large, e.g., thousands of bytes in each packet. However, the packets generated by MC-IoT are very small, e.g., $20$ or $32$ bytes in each packet \cite{3GPP2017Scenarios}. To achieve low latency for short packet transmissions, a short frame structure is adopted in 5G New Radio (NR) \cite{3GPP2017Agree}. When transmitting a short packet in a short frame, the blocklength of channel codes is very limited. As a result, the decoding error probability cannot be ignored when analyzing reliability \cite{Yury2010Channel}.

On the other hand, the computing ability at each MC-IoT device is limited. To reduce processing delay, MC-IoT devices will offload some of the packets to the Mobile Edge Computing (MEC) systems for processing \cite{Shih2017Fog,Yuyi2017MEC}. Considering that MC-IoT services will co-exist with eMBB services, a short packet arriving at the MEC after long packets has to wait in a queue if the packets are processed with a First-Come-First-Serve (FCFS) order. To avoid long queueing delay, other scheduling orders at MEC systems should be considered.

Furthermore, the reliability and delay not only depend on the resource management and scheduling order but also depend on the traffic load. Considering that the radio resources and computing capacity at each Access Point (AP) are limited, the user association and offloading policy should be optimized to balance traffic loads. The problems for optimizing user association and offloading policy are NP-hard in general \cite{dinh2017offloading}. Low-complexity solutions to the NP-hard problems are in urgent need for MC-IoT since complicated searching algorithms will lead to long computation delay \cite{View20175GPPP}.

\subsection{Related Works}
To transmit short packets with low latency, the blocklength of channel codes is short. In the short blocklength regime, Shannon's capacity is not applicable since it cannot characterize the decoding error probability \cite{Yury2010Channel}. Recently, the maximal achievable rate with given decoding error probability in the short blocklength regime was obtained in multi-antenna quasi-static channel \cite{Yury2014Quasi}. How to design transmission schemes and resource allocation in the short blocklength regime has been studied in existing literature \cite{Mustafa2013Throughput,Yulin2016Blocklength,Hu2018_NETWORK,Shengfeng2016Convexity,Cross2018she,she2017joint,Hu2018TWC_DL_QoS}. The throughput achieved in cognitive radio channels and relay systems was studied in \cite{Mustafa2013Throughput} and \cite{Yulin2016Blocklength,Hu2018_NETWORK}, respectively. The studies in \cite{Shengfeng2016Convexity} optimized the scheduling of short packets to maximize energy efficiency. The authors of \cite{Cross2018she} optimized packet losses caused by decoding errors, queueing delay violation, and packet dropping over deep fading channel subject to the ultra-high reliability. Considering that the feedback of Channel State Information (CSI) leads to extra delay, the studies in \cite{she2017joint} jointly optimized Uplink (UL) and Downlink (DL) resource configurations without CSI at the transmitters. More recently, how to optimize resource allocation among multiple users with different packet arrival processes was studied in \cite{Hu2018TWC_DL_QoS}.

Scheduling policies in computing systems have significant impacts on the Quality-of-Service (QoS) of MC-IoT. A near-optimal policy to minimize the average latency of short packet is the Shortest Remaining Processing Time (SRPT) first scheduler. Such a scheduler is hard to implement in practice since the remaining processing time is not available at the server, and it requires too many priority levels\cite{Homa2018}. To reduce the latency of short packets without introducing priority levels, the Processor-Sharing (PS) server is a possible solution, where the total service rate is equally allocated to all the packets in the server\cite{Mor2013Queue}. Although the distribution of latency was derived in the large delay regime in the PS server \cite{Sojourn2000Zwart}, the latency experienced by short packets remains unclear. To derive the delay bound violation probability of URLLC services, martingales-based analysis, effective capacity, and network calculus were used in \cite{zhao2018martingales}, \cite{Yulin2016Blocklength}, and \cite{Gross2015Delay}, respectively. But all the results were obtained in the FCFS servers. Note that it is very challenging to derive the closed-form expression of the distribution of delay, how to formulate the constraints on delay and reliability of MC-IoT is still unclear.

Promising network architectures for MC-IoT were studied in \cite{Emiliano2018IIoT,Xiaomin2018Adaptive,Nguyen2018Energy,Mohammad2019Fog,Zhong2018Traffic}. A comprehensive overview on MC-IoT of industrial scenarios was carried out in \cite{Emiliano2018IIoT}, where the issues related to architecture design were discussed, such as extensibility, scalability, and modularity. To reduce the routing delay, an adaptive transmission architecture with software-defined networks and edge computing was proposed in \cite{Xiaomin2018Adaptive}. Considering that energy consumption of IoT devices is an important issue, an energy-aware real-time routing scheme was proposed in \cite{Xiaomin2018Adaptive} to reduce energy consumption and E2E delay in large-scale IoT networks. More recently, a fog computing architecture was proposed for 5G tactile IoT \cite{Mohammad2019Fog}, where the quality-of-experience-aware model was formulated. By combining stochastic geometry and queueing theory, different QoS requirements were analyzed in ultra-dense networks \cite{Zhong2018Traffic}. The studies in \cite{Emiliano2018IIoT,Xiaomin2018Adaptive,Nguyen2018Energy,Mohammad2019Fog,Zhong2018Traffic} shed light upon network architecture designs for MC-IoT, but {decoding errors in the physical-layer were not considered.}

Computing offloading has been exhaustively studied in various MEC systems, such as wireless powered MEC \cite{Suzhi2018Computation}, ultra-dense IoT networks \cite{Hongzhi2018Mobile,Hongzhi2019Energy}, and fiber-wireless networks \cite{Bhaskar2017Cloudlet,Hongzhi2018Collaborative}. Although these works did not consider MC-IoT, they developed useful methodologies for optimizing computing offloading in MEC systems. How to improve the QoS in MEC systems by optimizing task offloading has been addressed in \cite{Juan2016MEC,Liu2017MEC,Jianhui2018Offloading,Seung2018MEC}. In \cite{Juan2016MEC}, the average delay was minimized by optimizing task offloading/scheduling in MEC systems. Considering that the average delay is not suitable for URLLC services, the authors of \cite{Liu2017MEC} optimized task offloading and resource allocation under the constraint on a queue length violation probability. The offloading schemes for URLLC in MEC systems were optimized in \cite{Jianhui2018Offloading}, where a weighted sum of E2E delay and the offloading failure probability was minimized in a single-user scenario. How to analyze latency in large-scale MEC networks was studied in \cite{Seung2018MEC}, where the average communication and computing latencies were derived.

Most of the existing studies on resource management in MEC systems only analyzed UL transmission and processing delay, and assumed DL transmission can be finished with high transmit power at the APs \cite{Juan2016MEC,Liu2017MEC,Seung2018MEC,Jianhui2018Offloading}. Besides, they did not take the decoding errors in the short blocklength regime into account, which is crucial for MC-IoT. Although the block error probability was considered in the optimization problem in \cite{Jianhui2018Offloading}, the radio resource management was not optimized and the packet losses due to the delay bound violation were not considered.

\subsection{Our Contributions}
To the best of the authors' knowledge, there is no resource allocation scheme or offloading scheme that can achieve the target E2E delay and overall packet loss probability for MC-IoT in MEC systems. Moreover, how to design scheduling policy and whether the FCFS server is suitable for MC-IoT were not addressed. In order to achieve ultra-high reliability and ultra-low E2E delay for MC-IoT in MEC systems, the following three issues will be addressed in this work: 1) \emph{How to design scheduling and queueing policies in MEC servers and local servers to achieve ultra-high reliability and ultra-low E2E delay for short packets?} 2) \emph{How to characterize the statistical QoS of short packets when there are both short and long packets in MEC systems?} 3) \emph{How to improve the fundamental tradeoff between delay and reliability by optimizing user association, packets offloading, and bandwidth allocation in MEC systems?} Our major contributions are summarized as follows,
\begin{itemize}
\item We establish a framework for minimizing overall packet loss probability subject to E2E delay requirement in MEC systems, where processing delay and UL and DL transmission delays are taken into account. PS servers are equipped at MEC systems, where the total service rate is equally allocated to all the packets in each server, and every packet receives service at all times. As such, short packets can bypass long packets and achieve low latency.

\item We derive a closed-form approximation of the Complementary Cumulative Distribution Function (CCDF) of the latency experienced by short packets in the PS server. The approximation is accurate when the number of Central Processing Unit (CPU) cycles needed to process a long packet is much larger than that needed to process a short packet.

\item We optimize the user association scheme, packet offloading rates, and bandwidth allocation for short packets in MEC systems. We propose an algorithm to solve a mixed integer problem and analyze the convergence and the complexity of it. Our analysis shows that the difference between the obtained solution and the global optimal solution only results from searching numbers of subcarriers in a continuous domain.
\end{itemize}

Furthermore, our simulation results validate the accuracy of the closed-form approximation. Numerical results indicate that when increasing the number of antennas at the AP with a fixed processing capacity, communication is the bottleneck of the reliability when the number of antennas is small, and computing is the bottleneck when the number of antennas is large. Only in a very small region (e.g., from $16$ to $18$ antennas), the packet loss probability in communications is comparable to the processing delay violation probability. This implies our algorithm converges to a near optimal solution in most of the cases.

The rest of the paper is organized as follows. Section II describes the system model. Section III analyzes delay and reliability. Section IV studies how to optimize the association scheme, packet offloading rates and bandwidth allocation. Section V extends the algorithm into more general scenarios. Section VI provides simulation and numerical results. Section VII concludes the paper.

\section{System Model}
\subsection{The MEC System}
\begin{figure}[htbp]
        \vspace{-0.1cm}
        \centering
        \begin{minipage}[t]{0.3\textwidth}
        \includegraphics[width=1\textwidth]{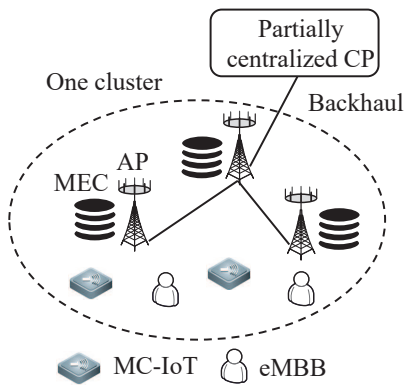}
        \end{minipage}
        \vspace{-0.2cm}
        \caption{System model.}
        \label{fig:system}
        \vspace{-0.2cm}
\end{figure}
As illustrated in Fig. \ref{fig:system}, we consider a MEC system with single-antenna devices and multi-antenna APs, where the data collected by each MC-IoT device and the computation intensive tasks generated by eMBB services can be offloaded to one of the APs for processing. To provide better services in radio access networks and to avoid high backhaul overhead, the partially centralized Control Plane (CP) in \cite{View20175GPPP} is considered. The whole network is decomposed into multiple clusters, each of which includes $K$ closely located APs and one CP that optimizes user association scheme, packet offloading rates, and bandwidth allocation for $M$ devices in the cluster. To achieve ultra-low latency and ultra-high reliability, strong co-channel interference should be avoided. To this end, orthogonal channels are allocated to different devices in each cluster, and the frequency reuse factor is less than one such that adjacent clusters use different bandwidth. In this work, we focus on one cluster of APs, and our solution is applicable for low mobility scenarios like factory automation and VR/AR applications. For high mobility scenarios, where devices travel across clusters frequently, how to reserve resources in different clusters deserves further study.

\subsection{Traffic Models}
In vehicle networks and factory automation, there are two kinds of packets, i.e., periodic packets with deterministic arrivals and sporadic packets that are driven by some random events \cite{Hassan2013A,Shehzad2016Emerg}. Since analyzing the delay and the reliability of random packet arrival processes is more challenging than deterministic arrivals, we focus on sporadic packets in this work. The experiment in \cite{hou2018Burstiness} indicates that the packet arrival processes of MC-IoT are very bursty, i.e., there is a high traffic state and a low traffic state. For each of the traffic states, the arrival process can be modeled as a Bernoulli process. According to 5G NR, time is discretized into slots with duration $T_{\rm s}$. In each slot, a device either has a packet to transmit or stays silent. We assume the traffic state is obtained with the traffic state classification methods in \cite{hou2018Burstiness}. When a device switches between the high and low traffic states, we only need to change the average arrival rate in our analysis. The aggregation of multiple independent Bernoulli processes at a MEC server can be accurately approximated by a Poisson process \cite{sriram1986characterizing}.

\subsubsection{Short packets} The data collected by each MC-IoT device is contained in \emph{short packets} for transmission and processing. A packet with the following three features is considered as ``short",
\begin{itemize}
\item The number of bits in the packet is small. According to \cite{3GPP2017Scenarios}, the packet size in MC-IoT services is around $20$ or $32$~bytes. In contrast, the packets in eMBB services may include hundreds or thousands of bytes, such as video streaming.

\item The blocklength of channel codes of the packet is short. For example, if quadrature phase-shift keying is used in modulation, the number of symbols required by a packet with $32$~bytes ($256$ bits) is $128$, which is the blocklength of the packet. To achieve low-latency, the blocklength of channel codes is short in MC-IoT \cite{Giuseppe2016Toward}.

\item The number of CPU cycles required to process the packet is small. The number of CPU cycles required to process the packet depends on the number of bits in the packet and the processing algorithms. Since a packet in MC-IoT services only contains a few bits, the number of required CPU cycles is small.
\end{itemize}
Let $c_{\rm S}$ be the number of CPU cycles required to process a short packet.  The service rate of the local server at the $k$th device and that of the $m$th MEC server are denoted as $C_k$ and $S_{m}$ $(\text{CPU cycles/slot})$, respectively.

\subsubsection{Long packets} There are some devices requesting eMBB services in each cluster. The tasks generated by the eMBB services are packetized into \emph{long packets}. How to optimize resource allocation and computing offloading for eMBB services has been studied in the existing literature, such as \cite{Kang2018Energy,Changsheng2018Asynchronous,Xianfu2018Performance}. In our work, we focus on MC-IoT services, where eMBB services are considered as background services.

The sum of average packet arrival rates of eMBB services at the $m$th AP is denoted as $\lambda^{\rm L}_m$ (packets/slot). The number of CPU cycles required to process a long packet is denoted as $c_{\rm L}$, which is a random variable with mean value $\bar{c}_{\rm L}$. In this work, we do not specify the distribution of $c_{\rm L}$. The only assumption on $c_{\rm L}$ is that $c_{\rm L} \gg c_{\rm S}$, which is reasonable since the packet size of eMBB services is much larger than that of MC-IoT services and the algorithm for processing long packets (e.g., high definition pictures) is more complex than that for processing short packets (e.g., the location and velocity of a device). For example, $c_{\rm L}$ may follow the Pareto distribution with a heavy tail \cite{Mor2013Queue}, i.e.,
\begin{align}\label{eq:CDFlong}
\Pr\left\{\frac{c_{\rm L}}{c_{\rm S}} > x\right\} = p_{\rm A} x^{-v},
\end{align}
where $1<v<2$, $p_{\rm A} = (c_0/c_{\rm S})^v$, and $c_0$ is the minimum of $c_{\rm L}$. As shown in \cite{Mor2013Queue} and the references therein, Pareto distributions have been observed in different application scenarios, such as the service time of UNIX jobs.

\subsection{User Association and Packet Offloading}
\subsubsection{User association} Each device can associate with one of the APs. We leverage indicators $x_{k,m}, k=1,...,K, m=1,...,M$, to represent the user association scheme,
\begin{align}
{x_{k,m}} = \left\{ {\begin{array}{*{20}{l}}
{1,{\text{ if device }}k{\text{ is associated with AP }}m{\rm{,}}}\\
{0,{\text{ otherwise}}{\rm{.}}}
\end{array}} \right.
\end{align}
The association scheme of the $k$th device is denoted as ${\bf{x}}_k = [x_{k,1},...,x_{k,M}]^{T}$.

\subsubsection{Packet offloading} When a short packet is generated by the $k$th device, the device either processes the packet with the local server or uploads the packet to an AP. The average packet rate from the $k$th device to the $m$th MEC server is denoted as $\lambda_{k,m}$ (packets/slot), where $k=1,2,...,K,$ and $m=0,1,...,M$. $m=0$ means that the packets are processed at the local server. If $x_{k,m'}=1$, then $\lambda_{k,0} + \lambda_{k,m'} = \lambda_k^{\rm U}$, where $\lambda^{\rm U}_k$ is the average packet arrival rate of the $k$th device. If $x_{k,m}=0$ for all $m=1,...,M$, then $\lambda_{k,0} = \lambda^{\rm U}_k$.

\subsection{Queueing Models and Scheduling Policies}
\begin{figure}[htbp]
        \vspace{-0.0cm}
        \centering
        \begin{minipage}[t]{0.25\textwidth}
        \includegraphics[width=1\textwidth]{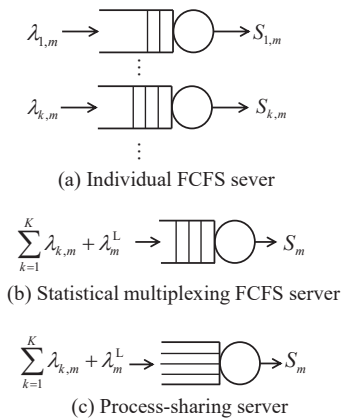}
        \end{minipage}
        \vspace{-0.2cm}
        \caption{MEC with different service orders.}
        \label{fig:queue}
        \vspace{-0.2cm}
\end{figure}

As shown in Fig. \ref{fig:queue}(a), to guarantee the QoS requirements of different devices, packets from different devices are waiting in different queues at a MEC server, and are served according to the FCFS order. In the $m$th MEC server, the service rate allocated to the $k$th device is denoted as $S_{k,m}$. Once the computing resource is allocated to one device, it cannot be shared with the other devices. Such a scheduling scheme is widely used, but is not optimal in terms of minimizing the delay.

The second server in Fig. \ref{fig:queue}(b) is referred to as a statistical multiplexing FCFS server \cite{Mor2013Queue}. Due to statistical multiplexing gain, the average delay in the second server is much shorter than the first server when $S_{m} =\sum\limits_k {{S_{k,m}}} $. Furthermore, as proved in \cite{she2017joint}, if the sizes of all the packets are identical, to achieve the same delay bound and delay bound violation probability, the required service rate in the statistical multiplexing server is less than the sum of the service rates in the individual server. However, when the distribution of the number of CPU cycles required to process the packets has a heavy-tail, some short packets arriving at the MEC server after a long packet need to wait for a long time. As a result, the delay requirement of MC-IoT services can hardly be satisfied.

The key to low latency is letting short packets bypass queued long packets. One possible solution is the PS server. As shown in Fig. \ref{fig:queue}(c), every packet in the server receives service at all times. When there are $i$ packets in the $m$th MEC server, each packet is processed at rate $S_{m}/i$.

\begin{rem}
\emph{In practice, a server can be implemented in a time-sharing way, i.e., the service time in each slot is equally allocated to all the packets in the server. In this way, the processing delay of packets in the server is the same as that in the ideal PS server \cite{Mor2013Queue}. It's worth noting that there are some other possible scheduling policies if the server is aware of the diverse QoS requirements of different packets. In this work, we do not assume the computing system is aware of the types of packets in the communication systems. We will study more sophisticated scheduling policies for different types of packets in our future work.}
\end{rem}

\begin{figure}[htbp]
        \vspace{-0.0cm}
        \centering
        \begin{minipage}[t]{0.3\textwidth}
        \includegraphics[width=1\textwidth]{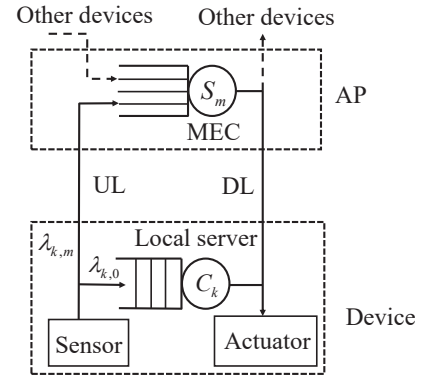}
        \end{minipage}
        \vspace{-0.2cm}
        \caption{Local and edge servers in our system.}
        \label{fig:OurMEC}
        \vspace{-0.2cm}
\end{figure}

We consider the scheduling policies at MEC servers and local servers in Fig. \ref{fig:OurMEC}, where PS servers are deployed at APs and FCFS servers are deployed at devices. To avoid queueing delay, the UL and DL transmission durations of a short packet equal to one slot.\footnote{Since the packet arrival rate of each device is less than one packet per slot, there is no queue before UL and DL transmissions.}  On the other hand, if the required CPU cycles to process different packets are identical, which is the case in the local server of each device, the FCFS server outperforms the PS server \cite{Mor2013Queue}. Therefore, FCFS servers are equipped at MC-IoT devices.

\section{Analysis of Delay and Reliability}
In this section, we study how to characterize E2E delay and overall packet loss probability. We first derive the CCDF of the processing delay of short packets in the PS server. Then, we show how to characterize the transmission delay and decoding error probability of short packets.

\subsection{Processing Delay and Delay Violation Probability}
A short packet can be processed either at the device or at the AP. The packet arrival process at the local server of the $k$th device is a Bernoulli process with average arrival rate $\lambda_{k,0}$. Denote the service time of a packet at the $k$th local server as $D_k^{\rm loc} = c_{\rm S}/C_k$. With the constant service rate at each local server, the queueing model is a Geo/D/1/FCFS model, where ``Geo" means that the inter-arrival time between packets is geometric distributed, and ``D" represents deterministic service processes. The CCDF of queueing delay in Geo/D/1/FCFS model has been obtained in \cite{Annie1990Performance}. If $i \le D_k^{\rm loc} - 1$, then
\begin{align}\label{eq:CCDFGeo}
\Pr\{D_k^{\rm q} > i\}={1-{{\left( {1 - {\lambda _{k,0}}} \right)}^{ - i-1}}\left(1-\lambda _{k,0}D_k^{\rm loc}\right).}
\end{align}
If $i \ge D_k^{\rm loc}$, the expression of $\Pr\{D_k^{\rm q} > i\}$ can be found in \cite{Annie1990Performance}.
Due to the low-latency requirement, we are interested in the case $i \le D_k^{\rm{loc}} - 1$.

Each AP may serve multiple devices. The aggregation of multiple Bernoulli processes can be modeled as a Poisson process \cite{Seung2018MEC}. Thus, the MEC server can be characterized by an M/G/1/PS model, where ``M" means the packet arrival process is Poisson process and ``G" means that the number of CPU cycles required to process the packets can follow any distributions. To derive a closed-form CCDF of the processing delay of short packets in the M/G/1/PS model, we introduce an accurate approximation. Since the short packets are much smaller than the long packets, i.e., $c_{\rm S} \ll c_{\rm L}$, the processing delay of a short packet is much shorter than a long packet. As a result, the number of long packets in the server is nearly constant from the arrival to the departure of a short packet. When a short packet arrives at the server, the number of packets in the server is denoted as $Q_{m}$. Considering that the value of $Q_{m}$ does not change significantly during the short service time of a short packet, then the service rate allocated to the short packet can be approximated by $S_{m}/(q+1)$ if $Q_{m} = q$. In this case, the processing delay of the short packet is approximated by
\begin{align}\label{eq:sojournQ}
W^{\rm S}_m|_{Q_{m} = q} \approx \frac{c_{\rm S}(q+1)}{S_m}\quad\text{(slots)}.
\end{align}

According to \cite{Mor2013Queue}, the distribution of $Q_m$ can be expressed as follows,
\begin{align}\label{eq:Qa}
\Pr\{Q_m = q\} = \rho_m^q(1-\rho_m),
\end{align}
which is the distribution of the number of packets in the M/G/1/PS model. The workload of the server is
\begin{align}\label{eq:LoadAll}
\rho_m = \frac{{\sum_{k=1}^{K}\lambda_{k,m}c_{\rm S}} + {\lambda^{\rm L}_m \bar{c}_{\rm L}}}{S_m}.
\end{align}
From \eqref{eq:sojournQ} and \eqref{eq:Qa}, we can further obtain that
\begin{align}
\Pr\left\{W^{\rm S}_m = \frac{c_{\rm S}(q+1)}{S_m} \right\} \approx  \Pr\{Q_m = q\} = \rho_m^q(1-\rho_m),\nonumber
\end{align}
where $q=0,1,...$. Based on the above expression, the CCDF of the processing delay of short packets in the PS server can be derived as follows,
\begin{align}
\Pr\left\{W^{\rm S}_m > \frac{c_{\rm S}(q+1)}{S_m} \right\} & \approx \Pr\{Q_m > q\} = \rho_m^q.\label{eq:CDFWs}
\end{align}
The approximation in \eqref{eq:CDFWs} is accurate when $c_{\rm S} \ll c_{\rm L}$. We will validate the accuracy of the approximation via simulation.

The processing delay of short packets in the $m$th MEC server can be characterized by a delay bound and a delay bound violation probability, $D^{\rm mec}_{m}$ and $\varepsilon^{\rm mec}_m$. From the CCDF in \eqref{eq:CDFWs}, the relationship between $\varepsilon^{\rm mec}_m$ and $D_{\rm m}^{\rm mec}$ can be expressed as follows,
\begin{align}
\varepsilon _m^{{\rm{mec}}} = \rho_m^{\left( {\frac{{{S_m}D_m^{{\rm{mec}}}}}{{{c_{\rm{S}}}}} - 1} \right)}. \label{eq:DSC}
\end{align}


\subsection{Transmission Delay and Decoding Error Probability}
If a packet is processed at the MEC server, the device first uploads the packet to the AP. After the MEC server finishes the processing, the result is sent back to the device. We introduce a superscript of parameters $X^{\xi}$, where $\xi\in \{{\rm{u}},{\rm{d}}\}$. If $\xi = \rm{u}$, $X$ is a parameter in UL transmissions. Otherwise, it is a parameter in DL transmissions. We consider Orthogonal Frequency Division Multiple Access (OFDMA) systems, which will be used to support MC-IoT services in 5G NR \cite{3GPP2017Agree}. The total bandwidth is equally allocated to $N_{\max}$ subcarriers, each with a bandwidth of $W_0$. Denote the number of subcarriers allocated to the $k$th device for UL and DL transmissions as $N^{\xi}_k$, $\xi \in \{{\rm{u}},{\rm{d}}\}$, respectively. Since the packet size is small, it is reasonable to assume that $N^{\xi}_kW_0$ is smaller than the coherence bandwidth. As mentioned in the previous section, to avoid queueing delay before UL and DL transmissions, the transmission duration of each packet is one slot, which is smaller than channel coherence time. Thus, each packet is transmitted over a flat fading quasi-static channel. Considering that feedback from receivers to transmitters may cause large overhead and extra delay, CSI is not available at the transmitters. According to \cite{Yury2014Quasi}, the achievable rate in the short blocklength regime over quasi-static flat fading channel can be accurately approximated by
\begin{align}
R^{\xi}_{k,m} \approx \frac{N^{\xi}_kW_0}{{\ln 2}}&\Bigg[ \ln \left( {1 + \frac{{{\alpha_{k,m}}{g^{\xi}_{k,m}}{P^{\xi}_{\rm s}}}}{{{N_0W_0}}}} \right) \nonumber\\
&- \sqrt {\frac{V_{k,m}^{\xi}}{{T_{\rm s}N^{\xi}_kW_0}}} f_{\rm{Q}}^{ - 1}\left( {{e^{\xi}_{{k,m}}}} \right) \Bigg]\;\text{bits/s},\label{eq:R}
\end{align}
where $\alpha_{k,m}$ is the large-scale channel gain from the $k$th device to the $m$th AP, $g^{\xi}_{k,m}$ is the UL or DL small-scale channel fading between the $k$th device and the $m$th AP, $P_{\rm s}^{\xi}$ is the UL or DL transmit power of one antenna on each subcarrier, $N_0$ is the single-side noise spectral density, $ f_{\rm{Q}}^{-1}(.)$ is the inverse of Q-function, ${e^{\xi}_{{k}}}$ is the decoding error probability, and $V_{k,m}^{\xi} = 1 - {1}\Big/{{{\left( 1 + \frac{{\alpha_{k,m}} {g^{\xi}_{k,m}}{P^{\xi}_{\rm s}}}{{{N_0W_0}}} \right)}^2}}$.

\begin{rem}
\emph{Due to the following two reasons, we only consider the flat fading channel, and do not consider frequency-selective channels. First, the maximal achievable rate over a frequency-selective channel in the short blocklength regime has not been derived in existing studies. Although the upper and lower bounds were obtained in \cite{Johan2017Low}, there is no closed-form expression. Second, as shown in \cite{she2017joint}, when the number of antennas is large (e.g., $16$ antennas), frequency diversity is not necessary for URLLC. Therefore, we focus on the multi-antenna flat fading channel. }
\end{rem}

Let $b^{\xi}_k$ be the number of bits in a short packet of the $k$th device. When sending a packet of $b^{\xi}_k$~bits within one slot, the decoding error probability can be obtained from \eqref{eq:R} by setting $T_{\rm s}R^{\xi}_{k,m} = b^{\xi}_k$. According to the law of total probability, the packet loss probability due to decoding errors can be expressed as follows \cite{Yury2014Quasi},
\begin{align}
{\varepsilon_{k,m}^{\xi}} &\approx {{\mathbb{E}}_{g^{\xi}_{k,m}}}\{{e^{\xi}_{{k,m}}}\}={{\mathbb{E}}_{g^{\xi}_{k,m}}}\Bigg\{ {f_{\rm Q}}\Bigg( \sqrt {\frac{T_{\rm s}N^{\xi}_kW_0}{V_k^{\xi}}} \nonumber\\
&\times\Bigg[ \ln \left( 1 + \frac{{{\alpha_{k,m}}{g^{\xi}_{k,m}}{P^{\xi}_{\rm s}}}}{{{N_0}W_0}} \right) - \frac{{b^{\xi}_k\ln 2}}{T_{\rm s}N^{\xi}_kW_0} \Bigg] \Bigg)\Bigg\},\label{eq:eu}
\end{align}
where the distribution of small-scale channel gain depends on the number of antennas at each AP, which is denoted as $N_{\rm t}$. To compute \eqref{eq:eu}, we need to calculate the integral for a given distribution of $g^{\xi}_{k,m}$. For Rayleigh fading, we can apply the closed-form result in \cite{She2018availability}.

\subsection{E2E Delay and Overall Packet Loss Probability}
\subsubsection{Delay and Reliability at Local Servers} If a packet is processed at the device, then the E2E delay is equal to the sum of the service time and the queueing delay at the local server of the device. Given the E2E delay requirement $D_{\max}$, the delay violation probability at local servers, $\varepsilon_k^{\rm loc}$, can be obtained by substituting $i = D_{\max}- D_k^{\rm{loc}}$ into \eqref{eq:CCDFGeo}. When $D_{\max}- D_k^{\rm{loc}} \le D_k^{\rm{loc}} - 1$,
\begin{align}
\varepsilon_k^{\rm loc} = 1-{{\left( {1 - {\lambda _{k,0}}} \right)}^{ - (D_{\max}- D_k^{\rm{loc}})-1}}\left(1-\lambda _{k,0}D_k^{\rm loc}\right).\label{eq:eUk}
\end{align}

\subsubsection{Delay and Reliability at the MEC server} If a packet is processed at a MEC server, then the UL and DL transmission delays and the processing delay in the server should be considered. The E2E delay can be satisfied under the following constraint,
\begin{align}
2 + D^{\rm mec}_{m} \leq D_{\max}, \label{eq:E2EDA}
\end{align}
where two slots are occupied by the UL and DL transmissions.

Due to decoding errors and processing delay violation, the overall packet loss probability can be expressed as follows,
\begin{align}
\varepsilon^{\rm A}_{k,m} &= 1-(1- \varepsilon_{k,m}^{\rm u})(1- \varepsilon_{k,m}^{\rm d})(1- \varepsilon _m^{{\rm{mec}}})\nonumber\\
&\approx \varepsilon_{k,m}^{\rm u} + \varepsilon_{k,m}^{\rm d} + \varepsilon _m^{{\rm{mec}}},\label{eq:E2EeA}
\end{align}
where the approximation is very accurate since $\varepsilon_{k,m}^{\rm u}$, $\varepsilon_{k,m}^{\rm d}$, and $\varepsilon _m^{{\rm{mec}}}$ are extremely small in MC-IoT. Upon substituting \eqref{eq:LoadAll} and \eqref{eq:E2EDA} into \eqref{eq:DSC}, we can get the expression of $\varepsilon_m^{{\rm{mec}}}$, i.e.,
\begin{align}
\varepsilon_m^{{\rm{mec}}} = \left(\frac{{\sum_{k=1}^{K}\lambda_{k,m}c_{\rm S}} + {\lambda^{\rm L}_m \bar{c}_{\rm L}}}{S_m}\right)^{ {\frac{{{S_m} (D_{\max}-2) }}{{{c_{\rm{S}}}}} - 1} }.\label{eq:ec}
\end{align}

\begin{rem}
\emph{The factors that lead to packet losses or errors depend on network architectures \cite{Daquan2019Toward}. For the considered MEC system, reliability only includes queueing delay violations in computing systems and packet losses in radio access networks.}
\end{rem}

\section{Cross-Layer Optimization}
In this section, we study how to optimize the association scheme, packet offloading rates, and UL and DL bandwidth allocation to minimize the overall packet loss probability subject to the E2E delay requirement.

\subsection{Problem Formulation}
Note that the packet loss probabilities at the local server and the MEC can be different, the reliability is determined by the worse one. Thus, the packet loss probability of the $k$th device is characterized by
\begin{align}
&f_k({\bf{x}}_k,\lambda_{k,m},N^{\rm u}_k,N^{\rm d}_k)\nonumber\\
&=\max\left[\varepsilon_k^{\rm loc},x_{k,m}(\varepsilon_{k,m}^{\rm u} + \varepsilon_{k,m}^{\rm d} + \varepsilon_m^{{\rm{mec}}}), \forall m\right].\label{eq:Lossk}
\end{align}
The problem that minimizes the maximal packet loss probability experienced by the $K$ devices can be formulated as follows,
\begin{align}
&\mathop {\mathop {\min }\limits_{{{\bf{x}}_k},{\lambda _{k,m}},N_k^{\rm{u}},N_k^{\rm{d}}} }\limits_{k = 1,...,K}\; \mathop {\max }\limits_{k = 1,...,K} {f_k}({{\bf{x}}_k},{\lambda _{k,m}},N_k^{\rm{u}},N_k^{\rm{d}})\label{eq:objLoss}\\
{\text{s.t.}}\; & \; \sum_{m=1}^M {x_{k,m}}\leq1, x_{k,m}\in \{0,1\},\label{eq:indicator} \tag{\theequation a}\\
&\; 0\leq \lambda_{k,m} \leq x_{k,m}, \label{eq:lamdba} \tag{\theequation b}\\
&\; \sum_{m=1}^M{\lambda_{k,m}} + \lambda_{k,0} = \lambda_{k}^{\rm U}, \label{eq:sumrate} \tag{\theequation c}\\
&\; \sum_{k=1}^K{N_k^{\rm{u}}} + \sum_{k=1}^K{N_k^{\rm{d}}} \leq N_{\max}, N_k^{\rm{u}}, N_k^{\rm{d}} \in \{1,2,...,N_{\rm c}\}, \label{eq:band} \tag{\theequation d}\\
&\; \max\{\varepsilon_k^{\rm{loc}}, x_{k,m}(\varepsilon_{k,m}^{\rm u}+ \varepsilon_{k,m}^{\rm d}+ \varepsilon_m^{{\rm{mec}}}), \forall m\} \leq 1, \label{eq:proba} \tag{\theequation e}
\end{align}
where $k = 1,...,K$, $m =1,...,M$, $N_{\rm c}$ is the maximum number of subcarriers that can be allocated to a device without exceeding the coherence bandwidth, and the expressions of $\varepsilon_{k,m}^{\xi}$, $\varepsilon_k^{\rm loc}$, and $\varepsilon_m^{{\rm{mec}}}$ can be found in \eqref{eq:eu}, \eqref{eq:eUk}, and \eqref{eq:ec}, respectively. Constraint \eqref{eq:indicator} guarantees that a device can only associate with one AP. If $\sum_{m=1}^M {x_{k,m}} = 0$, then all the packets are processed at the local server.

With constraint \eqref{eq:indicator}, each device cannot be served by two or more APs. Constraint \eqref{eq:lamdba} ensures that the packet offloading rate $\lambda_{k,m}$ is zero if the $k$th device is not served by the $m$th AP. Constraint \eqref{eq:sumrate} guarantees that the sum of the packet offloading rates at the APs and the packet arrival rate at the local server is equal to the total packet arrival rate of a device. The constraint on the maximal number of subcarriers of the system is given in \eqref{eq:band}, where the UL and DL bandwidth allocated to each device does not exceed the coherence bandwidth. When constraint \eqref{eq:proba} is satisfied, $\varepsilon_k^{\rm{loc}}$ and $\varepsilon_m^{{\rm{mec}}}$ are smaller than $1$, and hence the local and MEC servers are stable. By minimizing the objective function, we can check whether constraint \eqref{eq:proba} can be satisfied or not. If it cannot be satisfied, the problem is infeasible.

Since CSI is not available at the transmitters, the transmit power on each subcarrier is fixed. In UL transmission, the maximal transmit power of a device, $P^{\rm U}_{\max}$, is equally allocated to $N_{\rm c}$ subcarriers, $P_{\rm s}^{\rm u} = \frac{P^{\rm U}_{\max}}{N_{\rm c}}$. In DL transmission, the maximal transmit power of an AP, $P^{\rm A}_{\max}$, is equally allocated to $N_{\rm t}$ antennas. Considering that the number of subcarriers for DL transmission can be up to $N_{\max}$, to satisfy maximal transmit power constraint, the transmit power on each subcarrier is fixed as $P^{\rm A}_{\max}/N_{\max}$. Thus, we have $P_{\rm s}^{\rm d} = \frac{P^{\rm A}_{\max}}{N_{\max}N_{\rm t}}$.

Problem \eqref{eq:objLoss} is a mixed integer optimization problem, which is non-convex. In typical scenarios, the throughput of eMBB services is much higher than the throughput of MC-IoT services, and hence the number of CPU cycles required to process long packets are much larger than that required to process short packets. In the rest part of this section, we first consider the scenario that $({\sum_{k=1}^{K}\lambda^{\rm U}_{k}c_{\rm S}}) / ({\lambda^{\rm L}_m \bar{c}_{\rm L}})  \to 0$, and then extend our algorithm into more general scenarios.

\subsection{Solution in the Typical Scenario}
\subsubsection{Simplified Optimization Problem} According to \eqref{eq:sumrate}, $\lambda_{k,m} \leq \lambda^{\rm U}_{k}$. Thus, we have
\begin{align}
\rho_m = \frac{{\sum_{k=1}^{K}\lambda_{k,m}c_{\rm S}} + {\lambda^{\rm L}_m \bar{c}_{\rm L}}}{S_m} \leq \frac{{\sum_{k=1}^{K}\lambda^{\rm U}_{k}c_{\rm S}} + {\lambda^{\rm L}_m \bar{c}_{\rm L}}}{S_m}\triangleq \rho^{\rm ub}_m.\label{eq:rhoUB}
\end{align}
When $({\sum_{k=1}^{K}\lambda^{\rm U}_{k}c_{\rm S}}) / ({\lambda^{\rm L}_m \bar{c}_{\rm L}}) \to 0$, the equality in \eqref{eq:rhoUB} holds, and $\varepsilon_m^{{\rm{mec}}}$ in \eqref{eq:ec} is a constant that does not depend on packet offloading rates. Moreover, the packet loss probabilities due to decoding errors in UL and DL transmissions, $\varepsilon_{k,n}^{\rm u}$ and $ \varepsilon_{k,n}^{\rm d}$, do not change with packet offloading rates. Thus, the second term in $\max\left[\varepsilon_k^{\rm loc},x_{k,m}(\varepsilon_{k,n}^{\rm u} + \varepsilon_{k,n}^{\rm d} + \varepsilon_m^{{\rm{mec}}}), \forall m\right]$ does not depend on packet offloading rates. By setting $\lambda_{k,m} = \lambda^{\rm U}_{k}$, all the packets are offloaded to the MEC servers. Then, $\varepsilon_k^{\rm loc} = 0$ and
\begin{align}
f_k({\bf{x}}_k,\lambda_{k,m},N^{\rm u}_k,N^{\rm d}_k) = \mathop {\max }\limits_{m=1,...,M}{x_{k,m}(\varepsilon_{k,m}^{\rm u} + \varepsilon_{k,m}^{\rm d} + \varepsilon_m^{{\rm{mec}}})}.\label{eq:fklb}
\end{align}
With \eqref{eq:fklb}, problem \eqref{eq:objLoss} can be simplified as follows,
\begin{align}
\mathop {\mathop {\min }\limits_{{{\bf{x}}_k},N_k^{\rm{u}},N_k^{\rm{d}}} }\limits_{k = 1,...,K}\; \mathop {\mathop {\max }\limits_{k = 1,...,K}}\limits_{m = 1,...,M}&{x_{k,m}(\varepsilon_{k,m}^{\rm u} + \varepsilon_{k,m}^{\rm d} + \varepsilon_m^{{\rm{mec}}})}\label{eq:SIMobj}\\
\text{s.t.}\;&\; \eqref{eq:indicator}, \eqref{eq:band}, \;\text{and}\; \eqref{eq:proba}.\nonumber
\end{align}

\subsubsection{Packet Loss Balance Algorithm} Denote the optimal solution and the minimal packet loss probability of problem \eqref{eq:SIMobj} as $({{\tilde{\bf{x}}_k},\tilde{N}_k^{\rm{u}},\tilde{N}_k^{\rm{d}}})$ and $\tilde{\varepsilon}^{\rm A}$, respectively. In the following, we propose a binary search algorithm to find the optimal solution. The basic idea of the algorithm is to keep $\varepsilon_{k,m}^{\rm u} + \varepsilon_{k,m}^{\rm d} + \varepsilon_m^{{\rm{mec}}}, k=1,...,K,m=1,...,M,$ below a threshold $\varepsilon_{\rm th}$, and search the minimal $\varepsilon_{\rm th}$ in the regime $(0,\varepsilon_{\rm in}]$, where $\varepsilon_{\rm in} \leq 1$ is an initial upper bound of the overall packet loss probability. We refer to the algorithm as the Packet Loss Balance (PLB) Algorithm.

For a given threshold of overall packet loss probability $\varepsilon_{\rm th}$, we search for the optimal association scheme and subcarrier allocation that minimize the total number of subcarriers. If the minimum number of subcarriers exceeds $N_{\max}$, then $\tilde{\varepsilon}^{\rm A} > \varepsilon_{\rm th}$. Otherwise, $\tilde{\varepsilon}^{\rm A} \leq \varepsilon_{\rm th}$.

The problem that minimizes the total number of subcarriers can be expressed as follows,
\begin{align}
\mathop {\mathop {\min }\limits_{{{\bf{x}}_k},N_k^{\rm{u}},N_k^{\rm{d}}} }\limits_{k = 1,...,K}\;&\sum_{k=1}^K{N_k^{\rm{u}}} + \sum_{k=1}^K{N_k^{\rm{d}}}\label{eq:SumN}\\
\text{s.t.}\;&\;x_{k,m}(\varepsilon_{k,m}^{\rm u} + \varepsilon_{k,m}^{\rm d} + \varepsilon_m^{{\rm{mec}}})\leq \varepsilon_{\rm th},\label{eq:Eall} \tag{\theequation a}\\
&\;N_k^{\rm{u}}, N_k^{\rm{d}} \in \{1,2,...,N_{\rm c}\}, \;\text{and}\; \eqref{eq:indicator},\nonumber
\end{align}
where constraint \eqref{eq:proba} is removed since $\varepsilon_{\rm th}<1$. The above problem can be decoupled into $K$ problems since the association schemes and bandwidth allocation of different devices are independent.

Given that the $k$th device is served by the $m'$th MEC server, ${{x}}_{k,m'} = 1$, the required number of subcarriers can be found from the following problem,\footnote{By solving problem \eqref{eq:objMUk} with different $m' = 1,...,M$, we can obtain the optimal user association scheme and related bandwidth allocation that minimize $N_k^{\rm{u}}+N_k^{\rm{d}}$.}
\begin{align}
{\mathop {\min }\limits_{N_k^{\rm{u}},N_k^{\rm{d}}}}& \;N_k^{\rm{u}}+N_k^{\rm{d}}\label{eq:objMUk} \\
\text{s.t.}\;&\;\varepsilon_{k,m'}^{\rm u} + \varepsilon_{k,m'}^{\rm d} +\varepsilon_{m'}^{{\rm{mec}}} \leq \varepsilon_{\rm th},\label{eq:Eallk} \tag{\theequation a}\\
&\;N_k^{\rm{u}}, N_k^{\rm{d}} \in \{1,2,...,N_{\rm c}\}.\nonumber
\end{align}
To solve the inter programming problem, we first relax $N_k^{\rm{u}}$ and $ N_k^{\rm{d}}$ as continuous variables, and find the optimal subcarrier allocation. Then, we discretize the number of subcarriers used in UL and DL transmissions. Note that only the discretization
step will cause performance loss, which is minor as shown in \cite{She2015Tcom}.

To solve problem \eqref{eq:objMUk}, we need the following property of \eqref{eq:eu}.
\begin{pro}\label{P:epsilon}
\emph{The packet loss probabilities $\varepsilon_{k,m'}^{\rm u}$ and $\varepsilon_{k,m'}^{\rm d}$ in \eqref{eq:eu} are convex in $N_k^{\xi}$.}
\begin{proof}
See proof in Appendix \ref{App:1}.
\end{proof}
\end{pro}

Further considering that $\varepsilon_{m'}^{{\rm{mec}}}$ in \eqref{eq:ec} does not change with $N_k^{\xi}$, constraint \eqref{eq:Eallk} is convex. Therefore, problem \eqref{eq:objMUk} is a convex problem, and can be solved by techniques such as the interior-point method \cite{boyd}. The algorithm for solving problem \eqref{eq:SumN} is provided in Table \ref{T:subAlgorithm}, where $\left\lceil x \right\rceil $ is the minimum integer that is equal to or higher than $x$.

\renewcommand{\algorithmicrequire}{\textbf{Input:}}
\renewcommand{\algorithmicensure}{\textbf{Output:}}
\begin{table}[htb]\small
	\caption{Access scheme and bandwidth allocation}
	\label{T:subAlgorithm}
	\vspace{-0.6cm}
	\begin{tabular}{p{8.5cm}}
		\\\hline
	\end{tabular}
	\begin{algorithmic}[1]
		\REQUIRE Threshold of overall packet loss probability $\varepsilon_{\rm th}(i)$ (in the $i$th step of the binary search).
		\ENSURE Access scheme, ${{\bf{x}}_k}(i)$, and bandwidth allocation, ${N}_k^{\rm{u}}(i)$ and ${N}_k^{\rm{d}}(i)$ (optimal solution of problem \eqref{eq:SumN} in the $i$th step of the binary search).
		\STATE Set $k:=1$ and $m:=1$.
		\WHILE{$k\leq K$}
        \WHILE{$m\leq M$}
        \STATE Set ${x}_{k,m}(i):=1$.
        \STATE Relaxing ${N}_k^{\rm{u}}(i)$ and ${N}_k^{\rm{d}}(i)$ as continuous variables $\hat{N}_k^{\rm{u}}(m)$ and $\hat{N}_k^{\rm{d}}(m)$, respectively.
        \STATE Solve convex optimization problem \eqref{eq:objMUk}, and obtain $\hat{N}_k^{\rm{u}}(m)$ and $\hat{N}_k^{\rm{d}}(m)$.
        \STATE Discretize the numbers of subcarriers, $\hat{N}_k^{\rm{u}}(m):= \left\lceil \hat{N}_k^{\rm{u}}(m) \right\rceil $ and $\hat{N}_k^{\rm{d}}(m):= \left\lceil \hat{N}_k^{\rm{d}}(m) \right\rceil $.
        \STATE Set $\hat{N}^{\rm tot}_{k}(m):=\hat{N}_k^{\rm{u}}(m)+\hat{N}_k^{\rm{d}}(m)$.
        \ENDWHILE
        \STATE Set $m' := \arg \mathop {\min }\limits_m  \hat{N}^{\rm tot}_k(m)$.
        \STATE Set ${x}_{k,m'}(i):=1$ and ${x}_{k,m}(i):=0$, $\forall m \ne m'$.
        \STATE Set ${N}_k^{\rm{u}}(i):=\hat{N}_k^{\rm{u}}(m')$ and ${N}_k^{\rm{d}}(i) := \hat{N}_k^{\rm{d}}(m')$
        \ENDWHILE
		\RETURN ${{\bf{x}}_k}(i)$, ${N}_k^{\rm{u}}(i)$ and ${N}_k^{\rm{d}}(i)$, $k=1,...,K$.
	\end{algorithmic}
	\vspace{-0.2cm}
	\begin{tabular}{p{8.5cm}}
		\\
		\hline
	\end{tabular}
	\vspace{-0.2cm}
\end{table}

Note that problem \eqref{eq:SumN} could be infeasible if $\varepsilon_{\rm th}$ is too small. In this case, the minimal overall packet loss probability is higher than $\varepsilon_{\rm th}$. Based on the algorithm in Table \ref{T:subAlgorithm}, the PLB algorithm for solving problem \eqref{eq:SIMobj} is shown in Table \ref{T:Algorithm}.

\renewcommand{\algorithmicrequire}{\textbf{Input:}}
\renewcommand{\algorithmicensure}{\textbf{Output:}}
\begin{table}[htb]\small
	\caption{Packet Loss Balance Algorithm}
	\label{T:Algorithm}
	\vspace{-0.6cm}
	\begin{tabular}{p{8.5cm}}
		\\\hline
	\end{tabular}
	\begin{algorithmic}[1]
		\REQUIRE Total number of subcarriers, $N_{\max}$, the bandwidth of each subcarrier, $W_0$, coherence bandwidth, $W_0N_{\rm c}$, UL and DL transmit power on each subcarrier, $P_{\rm s}^{\rm u}$ and $P_{\rm s}^{\rm d}$, large-scale channel gains of devices, $\alpha_k$, the initial search area, $(0,\varepsilon_{\rm in})$, required accuracy of packet loss probability, $\Delta_{\varepsilon}$.
		\ENSURE Access scheme, ${\tilde{\bf{x}}_k}$, bandwidth allocation, $\tilde{N}_k^{\rm{u}}$ and $\tilde{N}_k^{\rm{d}}$, and packet loss probability $\tilde{\varepsilon}^{\rm A}$.
		\STATE Set $i:=1$, $\varepsilon^{\rm LB}(i):=0$, $\varepsilon^{\rm UB}(i):= \varepsilon_{\rm in}$, and $\varepsilon_{\rm th}(i) := (\varepsilon^{\rm LB}(i)+\varepsilon^{\rm UB}(i))/2$.
		\WHILE{$\varepsilon^{\rm UB}(i)-\varepsilon^{\rm LB}(i)> \Delta_{\varepsilon}$}
        \STATE Solve problem \eqref{eq:SumN} with the algorithm in Table \ref{T:subAlgorithm}, and obtain ${\bf{x}}_k(i)$, ${N}_k^{\rm{u}}(i)$ and ${N}_k^{\rm{d}}(i)$.
		\IF {$\sum_{k=1}^K\left[{N}_k^{\rm{u}}(i) + {N}_k^{\rm{d}}(i)\right] > N_{\max}$ or problem \eqref{eq:objMUk} is infeasible}
		\STATE Set $\varepsilon^{\rm LB}(i+1) := \varepsilon_{\rm th}(i)$ and $\varepsilon^{\rm UB}(i+1) := \varepsilon^{\rm UB}(i)$.
		\ELSE
		\STATE Set $\varepsilon^{\rm UB}(i+1) := \varepsilon_{\rm th}(i)$ and $\varepsilon^{\rm LB}(i+1) := \varepsilon^{\rm LB}(i)$.
		\ENDIF
		\STATE Set $\varepsilon_{\rm th}(i+1) := (\varepsilon^{\rm LB}(i+1)+\varepsilon^{\rm UB}(i+1))/2$.
        \STATE $i := i+1$.
		\ENDWHILE
		\STATE Set $\tilde{\varepsilon}^{\rm A} := \varepsilon_{\rm th}(i-1)$, ${\tilde{\bf{x}}_k} := p_k(i-1)$, $\tilde{N}_k^{\rm{u}} := {N}_k^{\rm{u}}(i-1)$, and $\tilde{N}_k^{\rm{d}}:= {N}_k^{\rm{d}}(i-1)$, $k=1,...,K$.
		\RETURN $\tilde{\varepsilon}^{\rm A}$, ${\tilde{\bf{x}}_k}$, $\tilde{N}_k^{\rm{u}}$, and $\tilde{N}_k^{\rm{d}}$, $k=1,...,K$.
	\end{algorithmic}
	\vspace{-0.2cm}
	\begin{tabular}{p{8.5cm}}
		\\
		\hline
	\end{tabular}
	\vspace{-0.2cm}
\end{table}

\subsubsection{Convergence of the PLB Algorithm}
To prove that the PLB algorithm converges to the minimal packet loss probability of problem \eqref{eq:SIMobj}, we first prove the following proposition,

\begin{prop}\label{P:conv}
\emph{The minimal packet loss probability ${\varepsilon}^{\rm A*} $ lies in the region $ (\varepsilon^{\rm LB}(i), \varepsilon^{\rm UB}(i)]$, $\forall i\in\{1,2,3,...\}$.}
\begin{proof}
See proof in Appendix \ref{App:2}.
\end{proof}
\end{prop}
According to Proposition \ref{P:conv}, the minimal packet loss probability lies in the region $ (\varepsilon^{\rm LB}(i), \varepsilon^{\rm UB}(i)]$. After $i$ steps of searching, the gap between ${\varepsilon}^{\rm A*}$ and the output of the PLB algorithm, $\tilde{\varepsilon}^{\rm A}$, is smaller than $0.5[\varepsilon^{\rm UB}(i)-\varepsilon^{\rm LB}(i)]$. In addition, with the binary search algorithm (i.e. from Line 1 to Line 11 in Table \ref{T:Algorithm}), the range of $(\varepsilon^{\rm LB}(i),\varepsilon^{\rm UB}(i)]$ decreases according to the following expression, $\varepsilon^{\rm UB}(i)-\varepsilon^{\rm LB}(i) = \varepsilon_{\rm in}/{2^i}$. When $i$ is large enough, $\tilde{\varepsilon}^{\rm A}$ approaches to ${\varepsilon}^{\rm A*}$.

The above proof holds when the performance loss caused by the discretization step in Line 7 of Table \ref{T:subAlgorithm} is negligible. Since the discretization step inevitably causes some performance loss, the related association scheme and bandwidth allocation are near optimal.

\subsubsection{Complexity of the PLB Algorithm}
With the PLB algorithm, we need to solve problem \eqref{eq:SumN} around $\log_2(\varepsilon_{\rm in}/\Delta_{\varepsilon})$ times. Problem \eqref{eq:SumN} is decoupled into $K$ single-device problem in \eqref{eq:objMUk}. With the algorithm in Table \ref{T:subAlgorithm}, the convex optimization problem in \eqref{eq:objMUk} is solved $KM$ times for $K$ devices with $M$ possible APs. The complexity of solving the convex optimization problem is denoted as $\Omega_0$, which is not high. Therefore, the complexity of the PLB algorithm is $\mathcal{O}\left(\log_2(\varepsilon_{\rm in}/\Delta_{\varepsilon})KM\Omega_0\right)$. Considering that a device will not associate with an AP that is very far from it, $M$ will not be very large. For example, a device can only be connected to one of the three or four APs with the highest large-scale channel gains. As a result, the complexity of the PLB algorithm increases linearly with the number of devices.

\section{Solution in General Scenarios}
To solve problem \eqref{eq:objLoss}, we extend the PLB algorithm into general scenarios without the assumption $({\sum_{k=1}^{K}\lambda^{\rm U}_{k}c_{\rm S}}) / ({\lambda^{\rm L}_m \bar{c}_{\rm L}})  \to 0$.

\subsection{Extended PLB Algorithm}
Although problem \eqref{eq:objLoss} cannot be simplified as problem \eqref{eq:SIMobj}, we can still use the algorithm in Table \ref{T:Algorithm}. The difference between the general scenarios and the scenario with the assumption $({\sum_{k=1}^{K}\lambda^{\rm U}_{k}c_{\rm S}}) / ({\lambda^{\rm L}_m \bar{c}_{\rm L}})  \to 0$ lies in Line 3 of the algorithm, where problem \eqref{eq:SumN} is obtained from \eqref{eq:SIMobj}. In general scenarios, given the threshold of overall packet loss probability, $\varepsilon_{\rm th}$, the optimization problem that minimizes the total number of subcarriers can be expressed as follows,
\begin{align}
&\mathop {\mathop {\min }\limits_{{{\bf{x}}_k},\lambda_{k,m},N_k^{\rm{u}},N_k^{\rm{d}}} }\limits_{k = 1,...,K}\;\sum_{k=1}^K{N_k^{\rm{u}}} + \sum_{k=1}^K{N_k^{\rm{d}}}\label{eq:GSumN}\\
\text{s.t.}\;&\;\max[\varepsilon_k^{\rm loc}, x_{k,m}(\varepsilon_{k,m}^{\rm u} + \varepsilon_{k,m}^{\rm d} + \varepsilon_m^{{\rm{mec}}}), \forall m ]\leq \varepsilon_{\rm th},\label{eq:Eth} \tag{\theequation a}\\
&\;N_k^{\rm{u}}, N_k^{\rm{d}} \in \{1,2,...,N_{\rm c}\}, \eqref{eq:indicator}, \eqref{eq:lamdba},  \;\text{and}\; \eqref{eq:sumrate}.\nonumber
\end{align}

From the expression of $\varepsilon_m^{\rm{mec}}$ in \eqref{eq:ec}, we can see that $\varepsilon_m^{\rm{mec}}$ increases with the packet offloading rate $\lambda_{k,m}$. Besides, the expressions of $\varepsilon_{k,m}^{\rm u} $ and $ \varepsilon_{k,m}^{\rm d}$ in \eqref{eq:eu} show that the required number of subcarriers increases as $\varepsilon_{k,m}^{\xi}$ decreases. Therefore, to satisfy constraint \eqref{eq:Eth}, the number of subcarriers increases with the packet offloading rate $\lambda_{k,m}$. To minimize the total number of subcarriers, the first step is minimizing packet offloading rates.

\emph{Step 1:} Optimize offloading rates. To minimize packet offloading rates, we find the maximal packets arrival rate at the local server, denoted as $\tilde{\lambda}_{k,0}(i)$. Note that the delay violation probability at each local server should satisfy $\varepsilon_k^{\rm loc} \leq \varepsilon_{\rm th}$, by substituting the expression of $\varepsilon_k^{\rm loc}$ in \eqref{eq:eUk} into the constraint, $\tilde{\lambda}_{k,0}(i)$ can be obtained via binary search. If $\tilde{\lambda}_{k,0}(i) > \lambda_k^{\rm U}$, all the packets are processed at the local server, $\lambda_{k,m}(i)=0$, $x_{k,m}(i) = 0$, $m=1,...,M$ and $N_k^{\rm u}=N_k^{\rm d}=0$. Otherwise, the packet offloading rate of the $k$th device is $\sum_{m=1}^M{\lambda_{k,m}(i)} = \lambda_k^{\rm U}-\tilde{\lambda}_{k,0}(i)$. As such, the constraint on the packet offloading rate in \eqref{eq:sumrate} can be expressed as follows,
\begin{align}
\sum_{m=1}^M{\lambda_{k,m}}  = \max[0,\lambda_{k}^{\rm U}-\tilde{\lambda}_{k,0}(i)], k = 1,...,K. \label{eq:sumoffload}
\end{align}
According to \eqref{eq:indicator} and \eqref{eq:lamdba}, each device only offload it's packets to one AP. Thus, the value of $\lambda_{k,m}$ is determined by $x_{k,m}$. If $x_{k,m}=1$, $\lambda_{k,m} = \max[0,\lambda_{k}^{\rm U}-\tilde{\lambda}_{k,0}(i)]$. Otherwise, $\lambda_{k,m} = 0$.

With the minimal packet offloading rates, problem \eqref{eq:GSumN} can be simplified as follows,
\begin{align}
\mathop {\mathop {\min }\limits_{{{\bf{x}}_k},N_k^{\rm{u}},N_k^{\rm{d}}} }\limits_{k = 1,...,K}\;&\sum_{k=1}^K{N_k^{\rm{u}}} + \sum_{k=1}^K{N_k^{\rm{d}}}\label{eq:GSumN2}\\
\text{s.t.}\;&\; x_{k,m}(\varepsilon_{k,m}^{\rm u} + \varepsilon_{k,m}^{\rm d} + \varepsilon_m^{{\rm{mec}}}) \leq \varepsilon_{\rm th}, \label{eq:Eth2} \tag{\theequation a}\\
&\;\lambda_{k,m} = x_{k,m}\max[0,\lambda_{k}^{\rm U}-\tilde{\lambda}_{k,0}(i)] \label{eq:sumoffload2}\tag{\theequation a}\\
&\;N_k^{\rm{u}}, N_k^{\rm{d}} \in \{1,2,...,N_{\rm c}\}, \;\text{and}\; \eqref{eq:indicator}.\nonumber
\end{align}
where $k=1,...,K$, and $ m=1,...,M$.
Different from problem \eqref{eq:SumN}, problem \eqref{eq:GSumN2} cannot be decoupled into $K$ subproblems. This is because the workloads of the APs depend on association schemes of all the devices. As a result, $\varepsilon_m^{\rm{mec}}$ changes with $x_{k,m}$. Changing the association scheme of one device will lead to different overall packet loss probabilities of all the other devices.

Based on this fact that the throughput of eMBB services is higher than MC-IoT services in most of the scenarios, we optimize the association scheme given the optimal bandwidth allocation obtained from problem \eqref{eq:SumN}, and then update bandwidth allocation according to the association scheme and related workloads at MEC servers.

\emph{Step 2:} Optimize the association scheme ${\bf{x}}_k(i)$. We set $N_k^{\rm u} $ and $ N_k^{\rm d}$ as the values that are obtained under the assumption $({\sum_{k=1}^{K}\lambda^{\rm U}_{k}c_{\rm S}}) / ({\lambda^{\rm L}_m \bar{c}_{\rm L}})  \to 0$, and compute $\varepsilon^{\rm u}_{k,m}+\varepsilon^{\rm d}_{k,m}$, $k=1,...,K$, $m = 1,...,M$. The initial workloads of the MEC servers are $\hat{\rho}_m = {\lambda^{\rm L}_m \bar{c}_{\rm L}}/S_m$, $m=1,...,M$. From \eqref{eq:DSC} we can obtain the initial delay bound violation probability $\hat{\varepsilon}_m^{{\rm{mec}}}$. Then, from the $1$st device to the $K$th device, each device selects one AP that can minimize $\varepsilon^{\rm u}_{k,m}+\varepsilon^{\rm d}_{k,m}+\hat{\varepsilon}_m^{{\rm{mec}}}$. Denote $\hat{m}_k$ as the AP that minimizes $\varepsilon _{k,m}^{\rm{u}} + \varepsilon _{k,m}^{\rm{d}} + \hat \varepsilon _m^{{\rm{mec}}}$. The $\hat{m}_k$th element of ${\bf{x}}_k(i)$ equals to one and $\lambda_{k,\hat{m}_k} = \max[0,\lambda_{k}^{\rm U}-\tilde{\lambda}_{k,0}(i)]$, $\lambda_{k,m} = 0$, $\forall m \ne \hat{m}_k$. After the $k$th device is associated with the $\hat{m}_k$th AP, the workload is updated according to $\hat{\rho}_m = ({\lambda_{1,m}c_{\rm S}+\lambda_{2,m}c_{\rm S}+...+\lambda_{k,m}c_{\rm S}+\lambda^{\rm L}_m \bar{c}_{\rm L}})/S_m$.

\emph{Step 3:} Update bandwidth allocation $N_k^{\rm u}$ and $N_k^{\rm d}$. Given ${\bf{x}}_k(i)$, we solve problem \eqref{eq:objMUk} for each device, and obtain the bandwidth allocation.

\begin{rem}
\emph{The packet offloading rates obtained in Step 1 and the bandwidth allocation obtained in Step 3 are optimal with given ${\bf{x}}_k(i)$. If we can obtain the optimal association scheme in Step 2, then we can obtain the optimal solution of problem \eqref{eq:GSumN}. However, the final workloads of the APs are not exactly the same as the initial values. Thus, ${\bf{x}}_k(i)$ obtained in Step 2 is not optimal. However, ${\bf{x}}_k(i)$ is near optimal if the association scheme of MC-IoT services has little impacts on the workloads of the APs. To provide more insights, we will prove that ${\bf{x}}_k(i)$ is optimal in two asymptotic cases in the sequel.}
\end{rem}

\subsection{Optimal Access Schemes in Two Asymptotic Cases}
In this subsection, we derive the optimal association scheme of problem \eqref{eq:objLoss} in the two asymptotic cases: communication or computing is the bottleneck of the overall packet loss probability. Whether communication or computing is the bottleneck depends on the number of antennas at each AP and the processing ability of the MEC server. In the next section, we will show the assumption that either communication or computing is the bottleneck is reasonable for various system parameters.

\subsubsection{Communication is the Bottleneck}
When the MEC servers have enough computing resources, the processing delay bound violation probability is much smaller than packet loss due to decoding errors, i.e., $\varepsilon_m^{\rm mec} \ll \varepsilon_{k,m}^{\xi}$. In this case, communication is the bottleneck of reliability. Denote ${\bf{x}}^{\rm comm}_k, k =1,...,K,$ as the association scheme that the $k$th device is served by the $m_k^*$th AP, which has the highest large-scale channel gain among all the APs, i.e., $\alpha_{k,m_k^*} > \alpha_{k,{m'_k}}, \forall {m'_k} \neq m^*_k$. The following proposition indicates that ${\bf{x}}^{\rm comm}_k$ is the optimal association scheme if $\sum_{m=1}^{M}{x_{k,m}} = 1$.

\begin{prop}
\emph{ If $\sum_{m=1}^{M}{x_{k,m}} = 1$, then for any solution of problem \eqref{eq:objLoss}, $({{{x}}_{k,m'_k}}=1, {\lambda_{k,m}},N_k^{\rm{u}},N_k^{\rm{d}})$, we can always find another solution, $({{{x}}_{k,m^*_k}}=1, {\lambda_{k,m}},N_k^{\rm{u}},N_k^{\rm{d}})$, that can achieve smaller packet loss probability.}
\begin{proof}
Since $\alpha_{k,m^*_k} \geq \alpha_{k,m'_k}$ and the decoding error probability in \eqref{eq:eu} decreases with the large-scale channel gain, we have $\varepsilon^{\xi}_{k,m^*_k} \leq \varepsilon^{\xi}_{k,m'_k}$. Therefore,
\begin{align}
&\max\limits_{k=1,...,K} \max\left[\varepsilon_k^{\rm loc},x_{k,m'_k}(\varepsilon_{k,m'_k}^{\rm u} + \varepsilon_{k,m'_k}^{\rm d}), \forall m\right] \nonumber\\
&\leq \max\limits_{k=1,...,K} \max\left[\varepsilon_k^{\rm loc},x_{k,m^*_k}(\varepsilon_{k,m^*_k}^{\rm u} + \varepsilon_{k,m^*_k}^{\rm d})\right].\nonumber
\end{align}
This completes the proof.
\end{proof}
\end{prop}

When all the packets of the $k$th device are processed at the local server, $\lambda_{k,0} = \lambda_k^{\rm U}$, if $\varepsilon_k^{\rm loc} \leq \varepsilon_{k,m^*}^{\rm u}+\varepsilon_{k,m^*}^{\rm d}$,  then $x_{k,m} = 0, \forall m$, and $N_k^{\rm u}=N_k^{\rm d}=0$. Otherwise, the device uploads some packets to the MEC servers to achieve better reliability, and hence $\sum_{m=1}^{M}{x_{k,m}} = 1$.

\subsubsection{Computing is the Bottleneck}
When there are very limited computing resources at the APs and no processing unit at devices, the packet loss probability due to decoding errors is much smaller than the processing delay violation probability, i.e., $\varepsilon_{k,m}^{\xi}  \ll \varepsilon^{\rm mec}_m$. In this case, all the packets are processed at the MEC servers, and the objective function of problem \eqref{eq:objLoss} can be simplified as follows,
\begin{align}
\mathop {\max }\limits_{k = 1,...,K} x_{k,m}\varepsilon_m^{\rm mec}\label{eq:objcom}.
\end{align}
Let ${\bf{x}}^{\rm comp}_k, k =1,...,K,$ be the optimal association scheme when computing is the bottleneck, respectively. To find the optimal solution, we denote the average packet arrival rate at the $m$th MEC server as $\lambda_m^{\rm A}$. Then, $\lambda_m^{\rm A} = \sum_{k=1}^{K}\lambda_{k,m}$ and
\begin{align}
\sum_{m=1}^{M}{\lambda_m^{\rm A}}=\sum_{m=1}^{M}{\sum_{k=1}^{K}{\lambda_{k,m}}}.\label{eq:Req}
\end{align}
Given the average arrival rate of each MEC server, the processing delay bound violation probability can be expressed as $\varepsilon_m^{\rm mec} = \rho_m^{\frac{S_m(D_{\max}-2)}{c_{\rm S}}-1}$, where $\rho_m = \frac{\lambda^{\rm A}_mc_{\rm S}+\lambda_m^{\rm L}{\bar{c}}_{\rm L}}{S_m}$.

\begin{figure}[htbp]
        \vspace{-0.2cm}
        \centering
        \begin{minipage}[t]{0.5\textwidth}
        \includegraphics[width=1\textwidth]{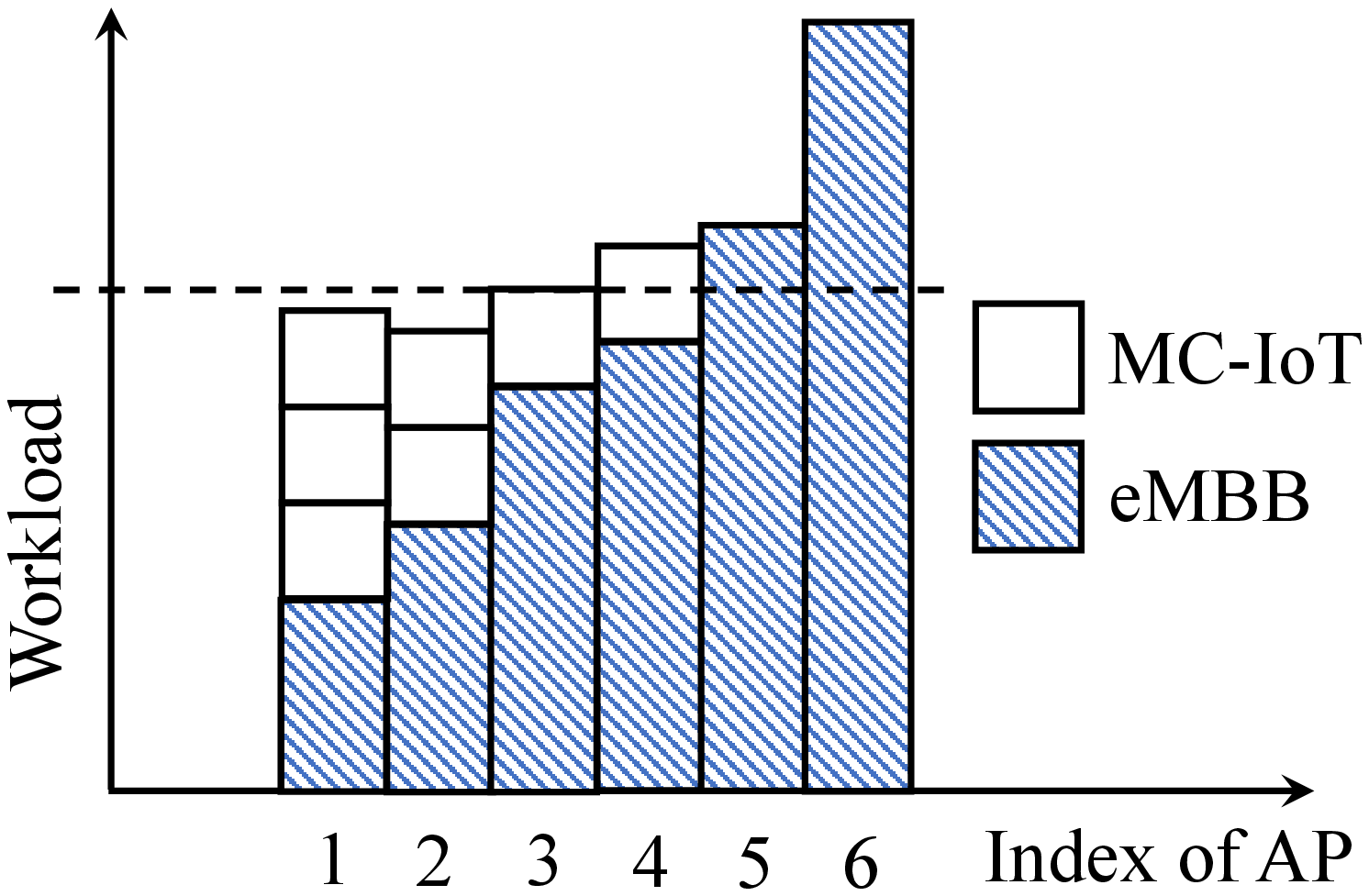}
        \end{minipage}
        \vspace{-0.2cm}
        \caption{Optimal association scheme when computing is the bottleneck.}
        \label{fig:CompBottle}
        \vspace{-0.2cm}
\end{figure}

Let $\varepsilon^{\rm A*}$ and $\rho^*$ be the minimal value of \eqref{eq:objcom} and the workload achieved by the optimal association scheme, respectively. If $\frac{\lambda_m^{\rm L}\bar{c}_{\rm L}}{S_m} \geq \rho^*$, then $x_{k,m} = 0, \forall k$, and $\lambda_m^{\rm A^*} = 0$. Let $\mathcal{M}$ be the set of MEC servers with $\frac{\lambda_m^{\rm L}\bar{c}_{\rm L}}{S_m} < \rho^*$. From $\rho^* = \frac{\lambda^{\rm A^*}c_{\rm S}+\lambda_m^{\rm L}\bar{c}_{\rm L}}{S_m}$, we can derive
\begin{align}
\lambda_m^{\rm A^*} = \frac{\rho^*S_m-\lambda_m^{\rm L}\bar{c}_{\rm L}}{c_{\rm S}}, \forall m \in \mathcal{M}.\label{eq:marrival}
\end{align}
Substituting $\lambda_m^{\rm A^*}$ into \eqref{eq:Req}, we can derive that
\begin{align}
\rho^* = \frac{{{\sum\limits_{{m=1}}^M{\sum\limits_{{k=1}}^{K}{\lambda_{k,m}}}{c_{\rm{S}}}}  + \sum\limits_{m \in \mathcal{M}} {\lambda _m^{\rm{L}}{{\bar c}_{\rm{L}}}} }}{{\sum\limits_{m \in \mathcal{M}} {{S_m}} }}\label{eq:optrho}.
\end{align}
To obtain $\rho^*$ and the related $\lambda_m^{\rm A^*}$, we need to obtain $\mathcal{M}$. Without loss of generality, we assume $\frac{\lambda_1^{\rm L}\bar{c}_{\rm L}}{S_1} \leq \frac{\lambda_2^{\rm L}\bar{c}_{\rm L}}{S_2}\leq...\leq \frac{\lambda_M^{\rm L}\bar{c}_{\rm L}}{S_M}$. Then, $\mathcal{M}$ can be expressed as $\mathcal{M} = \{m=1,..., M_{\rm th}\}$. Then, we can use the binary search algorithm to find the maximal $M_{\rm th}$ that satisfies $\lambda_m^{\rm A^*} > 0, \forall m \leq M_{\rm th}$. As illustrated in Fig. \ref{fig:CompBottle}, the basic idea of the optimal solution is offloading packets to the MEC servers with lower workloads.

On the one hand, since the association scheme is discretized, perhaps there is no association scheme that can keep the workloads of the $M_{\rm th}$ MEC servers exactly the same. On the other hand, if it is possible to satisfy $\rho_m = \rho^*$ in all the $M_{\rm th}$ servers, $x_k^{\rm comp}$ may not be unique. For example, if $\lambda^{\rm U}_{1} = \lambda_{2}^{\rm U}$, exchanging ${\bf{x}}^{\rm comp}_1$ and ${\bf{x}}^{\rm comp}_2$ does not change the workloads of the servers. Any association schemes that satisfy $\rho_m = \rho^*, \forall m \leq M_{\rm th},$ are optimal.

\subsection{Convergence of the Extended PLB Algorithm}
When communication is the bottleneck, $\varepsilon _{k,m}^{\rm{u}} + \varepsilon _{k,m}^{\rm{d}} + \hat \varepsilon _m^{{\rm{mec}}} \approx \varepsilon _{k,m}^{\rm{u}} + \varepsilon _{k,m}^{\rm{d}}$. Since the decoding error probability in \eqref{eq:eu} decreases with the large-scale channel gain, each device is associated with the MEC server with the largest large-scale channel gain. Thus, the association scheme obtained in Step 2 of the extended PLB algorithm is the same as the optimal association scheme when communication is the bottleneck, ${\bf{x}}_k^{\rm comm}$.

When computing is the bottleneck of reliability,  $\varepsilon _{k,m}^{\rm{u}} + \varepsilon _{k,m}^{\rm{d}} +  \varepsilon _m^{{\rm{mec}}} \approx  \varepsilon _m^{{\rm{mec}}}$. In the second step of the extended PLB algorithm, each device connected to the MEC server with the lowest workload. Then, the association scheme is the same as the optimal association scheme in Fig. \ref{fig:CompBottle}, ${\bf{x}}_k^{\rm comp}$.

\subsection{Complexity of the Extended PLB Algorithm} In the first step, we can use the binary search algorithm to find each $\lambda_{k,0}(i)$ with low complexity $\Omega_1$. In the second step, we find $\hat{m}_k$ from $M$ MEC servers for each of the $K$ devices. Denote the complexity for computing $\varepsilon _{k,m}^{\rm{u}} + \varepsilon _{k,m}^{\rm{d}} + \hat \varepsilon _m^{{\rm{mec}}}$ as $\Omega_2$. Then, the complexity of the second step is around $MK\Omega_2$. In the third step, we need to solve problem \eqref{eq:objMUk} $K$ times. Thus, the complexity is $K\Omega_0$. Therefore, the complexity of the extended PLB algorithm is $\mathcal{O}\left((K\Omega_0+K\Omega_1 + MK\Omega_2)\log_2(\varepsilon_{\rm in}/\Delta_{\varepsilon})\right)$, which is linear with the number of devices.

\section{Simulation and Numerical Results}
In this section, we validate the approximation of the CCDF of the processing delay of short packets in the PS server. To show the performance gain of our proposed framework, we compare the distributions of delay with PS servers and that with FCFS servers.\footnote{Although there are some related works, none of them took both uplink and downlink transmissions of short packets into account, and few of them optimized offloading policy from multiple devices to multiple APs with random packets arrival processes.} Besides, we illustrate the near optimal association scheme obtained with the extended PLB algorithm, and compared it with the optimal solution in the asymptotic cases. Finally, the reliability achieved by the extended PLB algorithm is illustrated in the scenarios with different communication and computing resources.

\subsection{Simulation Setup}
\begin{figure}[htbp]
        \vspace{-0.2cm}
        \centering
        \begin{minipage}[t]{0.5\textwidth}
        \includegraphics[width=1\textwidth]{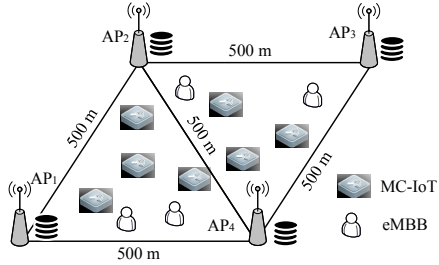}
        \end{minipage}
        \vspace{-0.2cm}
        \caption{Simulation Scenario}
        \label{fig:SS}
        \vspace{-0.4cm}
\end{figure}

The simulation scenario is shown in Fig. \ref{fig:SS}, where $4$ APs serve multiple MC-IoT and eMBB devices. The distance between two APs is $d_{\rm ap} = 500$~m. The path loss model is $35.3+37.6 \log_{10}\{d~(\text{m})\}$, where $d$ is the distance between an AP and a device served by it \cite{3GPP2011pathloss}. The shadowing is lognormal distributed with $8$~dB standard deviation. Half of the devices need $D_k^{\rm{s}} = 5$ slots to process a packet at their local servers, and the other half of the devices need $D_k^{\rm{s}} = 6$ slots. The packet arrival rate of each device, $\lambda_k^{\rm U}$, is uniformly distributed between $0.05$ and $0.1$ packets/slot {\cite{hou2018Burstiness}}.

The required CPU cycles for processing a long packet, $c_{\rm L}$, follows the distribution in \eqref{eq:CDFlong}.  Denote the processing delay of long packets at the $m$th MEC server as $W_m^{\rm L}$. According to the result in \cite{Sojourn2000Zwart}, we have
\begin{align}\label{eq:TailLong}
\Pr\{W_m^{\rm L} > x\} \sim p_{\rm A}(S_m/c_{\rm S})^{-v}(1-\rho_m)^{-v}x^{-v},
\end{align}
where $f(x)\sim h(x)$ means that $\lim_{x \to \infty} \frac{f(x)}{h(x)} = 1$. In our simulation, $v=1.5$.
Other parameters are listed in Table \ref{T:System}, unless otherwise specified.

\begin{table}[htbp]
\vspace{-0.2cm}\small
\renewcommand{\arraystretch}{1.3}
\caption{System Parameters \cite{3GPP2017Scenarios,3GPP2017Agree}}
\label{T:System}
\begin{center}\vspace{-0.2cm}
\begin{tabular}{|p{5.8cm}|p{2cm}|}
  \hline
  Transmit power of each device $P_{\rm tot}^{\rm M}$& $23$~dBm  \\\hline
  Transmit power of each AP $P_{\rm tot}^{\rm A}$& $46$~dBm  \\\hline
  Duration of each slot $T_{\rm s}$ & $0.125$~ms  \\\hline
  {E2E delay requirement $D_{\max}$} & {$1$~ms}  \\\hline
  Bandwidth of each subcarrier $W_{0}$& $120$~kHz \\\hline
  Number of subcarriers in each cluster $N_{\max}$& $256$ \\\hline
  Coherence bandwidth $N_{\rm c} W_{0}$& $1.2$~MHz \\\hline
  Packet size $b^{\xi}_k$ & $32$~bytes
  \\\hline
  The required minimal CPU cycles of long packets $c_0/c_{\rm S}$ & $10$ \\\hline
  The average arrival rate of long packets at each AP & $0.1$~packets/slot\\\hline
  Single-sided noise spectral density $N_0$ & $-174$~dBm/Hz \\\hline
\end{tabular}
\end{center}
\vspace{-0.6cm}
\end{table}

\subsection{CCDFs of Processing Delay}
\begin{figure}[htbp]
\vspace{-0.4cm}
\centering
\subfigure[{Long packets}]{
\label{fig:Long} 
\includegraphics[width=0.48\textwidth]{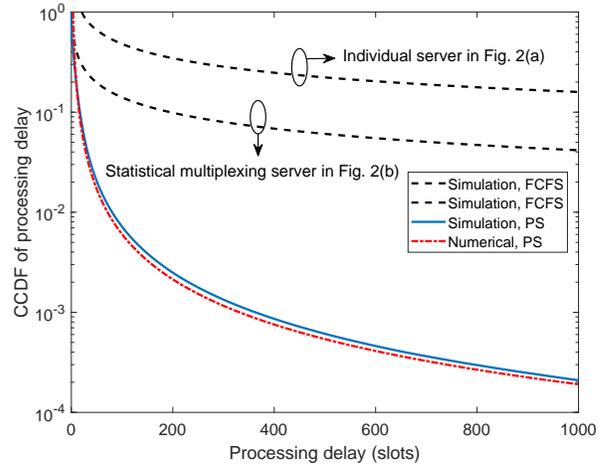}}
\vspace{-0.2cm}
\subfigure[{Short packets}]{
\label{fig:Short} 
\includegraphics[width=0.48\textwidth]{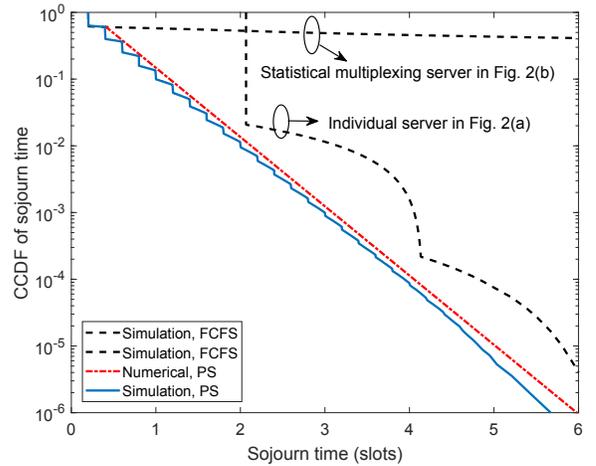}}
\caption{CCDFs of processing delay in the AP, where the service rate of the MEC is $S_m/c_{\rm S} = 5$ packets/slot.}
 \label{fig:CCDF} 
\vspace{-0.4cm}
\end{figure}

The CCDFs of processing delay in the FCFS servers and the PS server at the $m$th AP are illustrated in Fig. \ref{fig:CCDF}. In the simulation, $20$ devices are served by the AP, where the first $10$ of them send short packets and the other $10$ send long packets. The average packet rate from each device is $\lambda_{k,m} = 0.01$~packets/slot. The individual and statistical multiplexing servers are illustrated in Fig. \ref{fig:queue}(a) and Fig. \ref{fig:queue}(b), respectively. In the individual server, the service rate of the MEC server is allocated to the devices according to their packet arrival rate, i.e., $\sum_{k=1}^{20}{S_{k,m}}= S_m$, $\frac{\lambda_{k,m}c_{\rm S}}{S_{k,m}} = \rho_m$, $k=1,...,10$, and $\frac{\lambda_{k,m}\bar{c}_{\rm L}}{S_{k,m}} = \rho_m$, $k=11,...,20$. The simulation results are obtained by generating $10^8$ packets and computing their processing delay. The numerical results in Fig. \ref{fig:Long} and Fig. \ref{fig:Short} are obtained from \eqref{eq:TailLong} and \eqref{eq:CDFWs}, respectively.

The results in Fig. \ref{fig:Long} indicate that to achieve the same delay bound for long packets, the delay bound violation probability of PS server is much smaller than that of the FCFS servers. The results in Fig. \ref{fig:Short} indicate that the approximation in \eqref{eq:CDFWs} is very accurate for short packets, and the PS server outperforms the FCFS servers in the short delay regime. Besides, we can see that compared with the statistical multiplexing FCFS server, the individual FCFS server achieves better QoS for short packets by sacrificing the QoS of long packets. However, the PS server can achieve better QoS than the FCFS servers for both long and short packets when the distribution of the number of CPU cycles required to process the packets has a heavy tail.

\subsection{Overall Packet Loss Probabilities}
\begin{figure}[htbp]
\vspace{-0.4cm}
\centering
\subfigure[{Overall packet loss probability v.s. $N_{\rm t}$, where $W_0 = 120$~kHz and $T_{\rm s} = 0.125$.}]{
\label{fig:Ts1} 
\includegraphics[width=0.48\textwidth]{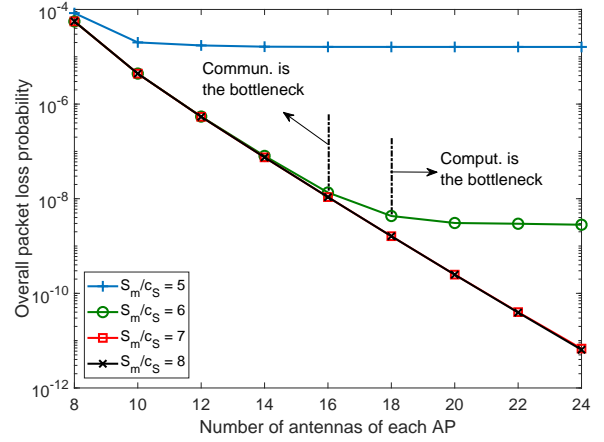}}
\vspace{-0.2cm}
\subfigure[{Overall packet loss probability v.s. $N_{\rm t}$ with different $W_0$ and $T_{\rm s}$.}]{
\label{fig:W0} 
\includegraphics[width=0.48\textwidth]{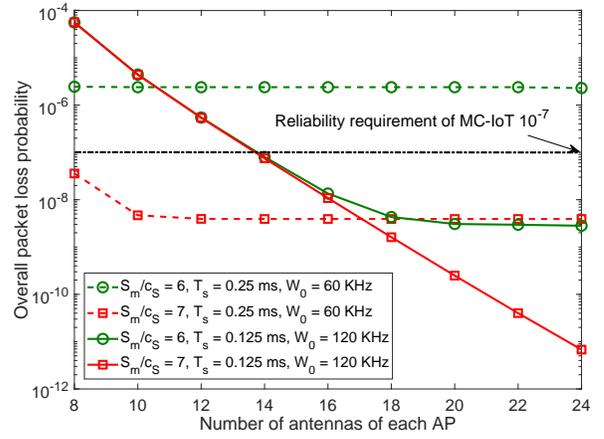}}
\caption{Overall packet loss probabilities achieved by the optimized association scheme, packet offloading rates, and bandwidth allocation, where $K=20$.}
 \label{fig:Reliability} 
\vspace{-0.2cm}
\end{figure}

The overall packet loss probabilities achieved by the optimized association scheme, packet offloading rates, and bandwidth allocation are illustrated in Fig. \ref{fig:Reliability}, where the solution and the minimal overall packet loss probability are obtained with the extended PLB algorithm. As shown in Fig. \ref{fig:Ts1}, better reliability can be achieved with more antennas or with higher service rate at the MEC servers. To show when communication or computing is the bottleneck of reliability, we consider the case $S_m/c_{\rm S} = 6$. When $N_{\rm t} \leq 16$, increasing the service rate of the MEC servers does not help improving reliability, and hence communication is the bottleneck. When $N_{\rm t} \geq 18$, overall packet loss probability does not decrease with $N_{\rm t}$, and hence computing is the bottleneck. As a result, only when $N_{\rm t}\in [16,18]$, the packet loss in UL and DL transmissions are comparable to the processing delay violation. In the cases that $S_m/c_{\rm S} > 6$, communication is always the bottleneck, because the processing delay violation probability is much smaller than the packet loss due to decoding errors. These results imply that the extended PLB algorithm converges to the optimal solutions in most of the scenarios.

According to 5G NR in \cite{3GPP2017Agree}, the bandwidth of each subcarrier, $W_0$, and the duration of each slot, $T_{\rm s}$, can be adjusted according to the requirements of services. In Fig. \ref{fig:W0}, we illustrated the impact of $W_0$ and $T_{\rm s}$ on the reliability, where $T_{\rm s}W_0$ is fixed as a constant such that there are $14$ symbols transmitted in each slot with one subcarrier. The results in Fig. \ref{fig:W0} indicate that when $N_{\rm t}$ is small (i.e., communication is the bottleneck), increasing $T_{\rm s}$ is helpful for increasing reliability. The total bandwidth allocated to each device does not exceed coherence bandwidth $1.2$~MHz, which does not change with $W_0$. By increasing $T_{\rm s}$, the maximal blocklength of each packet increases. As a result, the packet loss probability due to decoding errors decreases with $T_{\rm s}$. However, when computing is the bottleneck, increasing $T_{\rm s}$ leads to higher overall packet loss probability.

\begin{figure}[htbp]
        \vspace{-0.2cm}
        \centering
        \begin{minipage}[t]{0.48\textwidth}
        \includegraphics[width=1\textwidth]{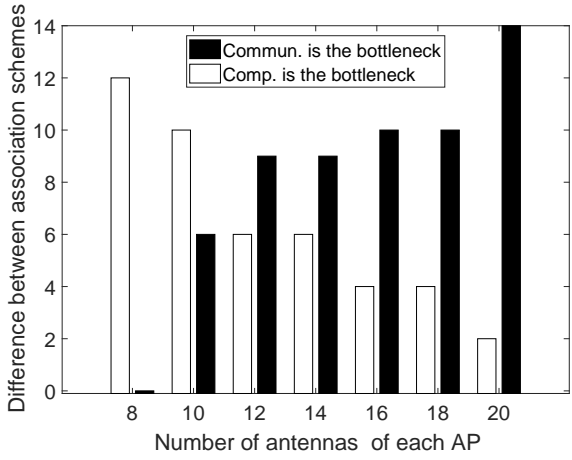}
        \end{minipage}
        \vspace{-0.2cm}
        \caption{The difference between the association schemes, where $S_m/c_{\rm S} = 6$ packets/slot and $K=20$.}
        \label{fig:DX}
        \vspace{-0.2cm}
\end{figure}
The differences between the association scheme obtained with the extended PLB algorithm and the optimal association schemes ( i.e., $ ||{\bf{x}}^{\rm PLB}_k-{\bf{x}}_k^{\rm comm}||$ with the legend ``Commun. is the bottleneck" and $||{\bf{x}}^{\rm PLB}_k-{\bf{x}}_k^{\rm comp}||$ with the legend ``Comp. is the bottleneck") are shown in Fig. \ref{fig:DX}. When $N_{\rm t}=8$, communication is the bottleneck and ${\bf{x}}^{\rm PLB}_k$ and ${\bf{x}}_k^{\rm comm}$ are the same. When $N_{\rm t}$ is large, ${\bf{x}}^{\rm PLB}_k$ approaches to ${\bf{x}}_k^{\rm comp}$. The results in Fig. \ref{fig:DX} are consistent with our analysis in Section V.C that the extended PLB algorithm converges to the optimal solutions in the two asymptotic cases.

\begin{figure}[htbp]
        \vspace{-0.2cm}
        \centering
        \begin{minipage}[t]{0.48\textwidth}
        \includegraphics[width=1\textwidth]{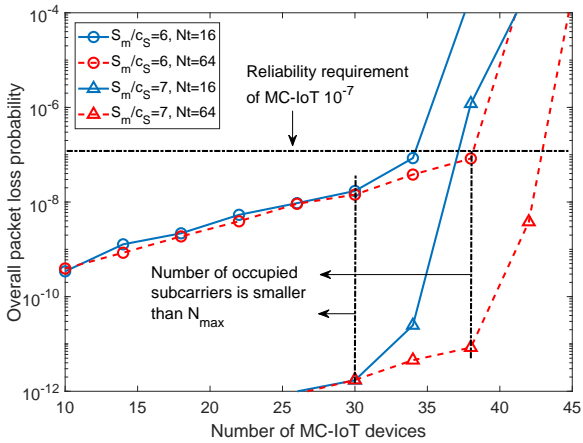}
        \end{minipage}
        \vspace{-0.2cm}
        \caption{Overall packet loss probability v.s. the number of MC-IoT devices in one cluster.}
        \label{fig:RvsK}
        \vspace{-0.2cm}
\end{figure}

The relation between the overall packet loss probability and the density of devices is illustrated in Fig. \ref{fig:RvsK}. The curves are not smooth, because problem \eqref{eq:objLoss} is a mixed integer optimization problem. The results in Fig. \ref{fig:RvsK} indicate that there is a tradeoff between the overall packet loss probability and the density of devices. When the number of occupied subcarriers is less than the maximal number of subcarriers, i.e., $\sum_{k=1}^{K}(N_k^{\rm u}+N_k^{\rm d}) < N_{\max}$, the overall packet loss probability increases slowly as $K$ increases. When $\sum_{k=1}^{K}(N_k^{\rm u}+N_k^{\rm d}) = N_{\max}$, the overall packet loss probability increases extremely fast as $K$ increases. By increasing $N_{\rm t}$ and $S_m$, the density of devices can be improved (i.e., the number of devices with overall packet loss probability less than $10^{-7}$), but the performance gain in Fig. \ref{fig:RvsK} is only around $25$\%.

%

\section{Conclusion}
In this work, we analyzed the processing delay of short packets in the M/G/1/PS server. By introducing an accurate approximation, the CCDF of the processing delay of short packets was derived in closed-form. We then formulated an optimization problem that minimizes the overall packet loss probability under the constraints on ultra-low E2E delay in the MEC system, where the association scheme, packet offloading rates, and bandwidth allocation for MC-IoT services are optimized. The problem is a mixed integer problem and is non-convex. To solve the problem, we proposed a PLB algorithm in the scenario that the data rates of eMBB services are much higher than that of MC-IoT services. We further extended the algorithm into more general scenarios, where we can obtain a near optimal solution with low complexity, i.e., the complexity increases linearly with the number of devices. To analyze the performance of the extended PLB algorithm, we derived the optimal solutions of the problem in two asymptotic cases: communication or computing is the bottleneck of reliability, and proved that the extended PLB algorithm converges to the optimal solution in these two asymptotic cases. Simulation and numerical results validated our analysis and showed that the PS server outperforms FCFS servers.


\appendices
\section{Proof of Property \ref{P:epsilon}}
\label{App:1}
\renewcommand{\theequation}{A.\arabic{equation}}
\setcounter{equation}{0}
\begin{proof}
Denote $f_N(N_k^{\xi}) = \sqrt {\frac{T_{\rm s}N^{\xi}_kW_0}{V_k^{\xi}}} \left[ \ln \left( 1 + \frac{{{\alpha_{k,m}}{g^{\xi}_{k,m}}{P^{\xi}_{\rm s}}}}{{{N_0}W_0}} \right) - \frac{{b^{\xi}_k\ln 2}}{T_{\rm s}N^{\xi}_kW_0} \right]$. The second order derivative of $f_N(N_k^{\xi})$ can be derived as follows,
\begin{align}
f''_N(N_k^{\xi})= &- \frac{1}{4}{\left( {N_k^\xi } \right)^{ - \frac{3}{2}}}\sqrt {\frac{{{T_{\rm{s}}}{W_0}}}{{V_k^\xi }}} \ln \left( {1 + \frac{{{\alpha _{k,m}}g_{k,m}^\xi P_{\rm{s}}^\xi }}{{{N_0}{W_0}}}} \right) \nonumber\\
&- \frac{3}{4}{\left( {N_k^\xi } \right)^{ - \frac{5}{2}}} {\frac{{b_k^\xi \ln 2}}{{\sqrt {{T_{\rm{s}}}{W_0}V_k^\xi } }}}  < 0 \label{eq:fn}.
\end{align}
Thus, $f_N(N_k^{\xi})$ is concave in $N_k^{\xi}$. Moreover, Q function $f_{\rm Q}(x)$ is a convex and decreasing function when $f_{\rm Q}(x) < 0.5$, which is the case in MC-IoT. According to \cite{boyd}, the composite function $f_{\rm Q}\left(f_N(N_k^{\xi})\right)$ is convex in $N_k^{\xi}$. This completes the proof.
\end{proof}

\section{Proof of Proposition \ref{P:conv}}
\label{App:2}
\renewcommand{\theequation}{B.\arabic{equation}}
\setcounter{equation}{0}
\begin{proof}
We apply the mathematical induction to prove Proposition \ref{P:conv}.
When $i=1$, $\varepsilon^{\rm LB}(1)=0$ and $\varepsilon^{\rm UB}(1) = \varepsilon_{\rm in}$, we have $\tilde{\varepsilon}^{\rm A} \in (\varepsilon^{\rm LB}(1), \varepsilon^{\rm UB}(1)]$. We assume that when $i=j$, $\tilde{\varepsilon}^{\rm A} \in (\varepsilon^{\rm LB}(j), \varepsilon^{\rm UB}(j)]$, and prove $\tilde{\varepsilon}^{\rm A} \in (\varepsilon^{\rm LB}(j+1), \varepsilon^{\rm UB}(j+1)]$.

In the case that $\sum_{k=1}^K\left[{N}_k^{\rm{u}}(i) + {N}_k^{\rm{d}}(i)\right] \leq N_{\max}$, $\varepsilon_{\rm th}(i)$ can be achieved by a solution that lies in the feasible region of problem \eqref{eq:SIMobj}. Thus, $\tilde{\varepsilon}^{\rm A} \leq \varepsilon_{\rm th}(i)$. According to the algorithm in Table \ref{T:Algorithm}, we have $\varepsilon^{\rm UB}(j+1) = \varepsilon_{\rm th}(j)$ and $\varepsilon^{\rm LB}(j+1) = \varepsilon^{\rm LB}(j)$. Further considering that $\tilde{\varepsilon}^{\rm A} > \varepsilon^{\rm LB}(j)$ with the assumption in the case $i=j$, we have $\tilde{\varepsilon}^{\rm A} \in (\varepsilon^{\rm LB}(j+1), \varepsilon^{\rm UB}(j+1)]$.

In the case that $\sum_{k=1}^K\left[{N}_k^{\rm{u}}(i) + {N}_k^{\rm{d}}(i)\right] > N_{\max}$, $\varepsilon_{\rm th}(i)$ cannot be achieved with $N_{\max}$ subcarriers. Thus, $\tilde{\varepsilon}^{\rm A} > \varepsilon_{\rm th}(i)$. According to the algorithm in Table \ref{T:Algorithm}, we have $\varepsilon^{\rm LB}(j+1) = \varepsilon_{\rm th}(j)$ and $\varepsilon^{\rm UB}(j+1) = \varepsilon^{\rm UB}(j)$. Further considering that $\tilde{\varepsilon}^{\rm A} \leq \varepsilon^{\rm UB}(j)$ with the assumption in the case $i=j$, we have $\tilde{\varepsilon}^{\rm A} \in (\varepsilon^{\rm LB}(j+1), \varepsilon^{\rm UB}(j+1)]$. The proof ends here.
\end{proof}

\bibliographystyle{IEEEtran}
\bibliography{ref_conference}

\end{document}